\documentclass[10pt, prb, aps, twocolumn,  citeautoscript, floatfix, reprint, amsmath, amssymb, notitlepage, superscriptaddress, longbibliography]{revtex4-2}

\usepackage[utf8]{inputenc}

\usepackage{amsmath,amssymb,amsfonts}
\usepackage{graphicx}
\usepackage{physics}
\usepackage{epsfig}
\usepackage{color}
\usepackage{mathrsfs}
\usepackage{txfonts}
\usepackage{xcolor}
\usepackage{subfigure}
\usepackage{comment}

\usepackage[breaklinks,colorlinks=true,linkcolor=blue,citecolor=blue,filecolor=blue,urlcolor=blue]{hyperref}

\DeclareUnicodeCharacter{2013}{-} 
\DeclareUnicodeCharacter{00A0}{ }
\DeclareUnicodeCharacter{0308}{HERE!}

\newcommand{\bea}{\begin{eqnarray}}
\newcommand{\eea}{\end{eqnarray}}
\newcommand{\dket}[1]{|#1\rangle}
\newcommand{\dbra}[1]{\langle#1|}

\newcommand{\bpm}{\begin{pmatrix}}
\newcommand{\epm}{\end{pmatrix}}

\begin{document}
	\title{Phase diagram of amorphous quantum spin Hall insulators}
	\author{Ranadeep Roy}
	\author{Yuan-Ming Lu}
	\affiliation{Department of Physics, The Ohio State University, Columbus OH 43210, USA}
	\date{\today}
	
	\begin{abstract}
In light of recent progress in the study of amorphous topological phases, we investigate the effects of structural disorder on the topological properties of a two-dimensional quantum spin Hall insulator modeled by the Bernevig-Hughes-Zhang Hamiltonian. Using a real-space formulation of the $\mathbb{Z}_2$ invariant for Dirac-type Hamiltonian, we map out the phase diagram as a function of disorder strength and the mass parameter. Our results reveal that under the influence of structural disorder, a system can either undergo a phase transition from a topologically non-trivial to a topologically trivial (N$\rightarrow$T) phase or from a trivial to non-trivial phase (T$\rightarrow$N). Remarkably, in certain parameter regimes, the system exhibits a re-entrant behaviour: a topologically non-trivial phase in the perfect lattice undergoes a transition to a trivial state under the influence of weak disorder but re-emerges as the disorder strength is further increased (N$\rightarrow$T$\rightarrow$N). We corroborate these findings through analysis of the bulk-boundary correspondence and transport calculations.
\end{abstract}
\maketitle

\section{Introduction}

Topological phases of matter have been the subject of intense study over the past two decades, driven by their novel physical properties and potential technological applications\cite{hasan,qi}. A defining feature of these systems is that they possess an energy gap in the bulk while supporting robust gapless excitations at their boundaries. These edge states are protected by the bulk topology of the system and remain stable in the presence of weak disorders\cite{schnyder,Chiu2016}. The interplay between disorder and topology often leads to rich and counterintuitive phenomena, a prominent example of which is the topological Anderson insulator\cite{Li1}, where strong onsite disorder can drive a trivial system into a topological phase. More recently, the seminal works of Mitchell et al.\cite{mitchell} and Agarwala and Shenoy\cite{agarwala1}, demonstrated that even a completely random lattice can support topological phases, providing an impetus to search for physical realizations of topological amorphous matters. Since then, several studies have explored the consequence of the lack of crystalline symmetries on the topological properties in a wide range of physical systems, both theoretically \cite{agarwala1,alvarez,regis,cheng,fleury1,agarwala2,varjas,grushin,corbae-review,marsal, Cassella2023-nq,yang,mansha, costa, spring1,kim,manna,martinez2025,ghosh2025,jezequel2025} and experimentally\cite{Corbae2023-kq,Ciocys2024,zhou2020,rechtsman,stutzer}.  

Among the myriad of known topological phases, an important type of topological phase is the $\mathbb{Z}_2$ topological insulator, which respects time-reversal symmetry and is characterized by a $\mathbb{Z}_2$ valued invariant. Previously, the effects of Anderson disorders\cite{Li1,groth2,orth2016} and random bond disorders\cite{song1} have been studied in such topological insulators. In this work, we numerically study the effects of amorphization on the topological properties of the Bernevig-Hughes-Zhang (BHZ) model \cite{bernevig-science} - a prototypical model of time-reversal symmetric, two-dimensional topological insulator, with extended hoppings. To continuously tune the system from a crystalline to an amorphous configuration, we introduce structural disorder via random displacements of the lattice sites, drawn from a normal distribution. We diagnose the topological nature of the system by using a real space $Z_2$ marker\cite{wchen1,oliveira}, tailored to Dirac-type Hamiltonians. Our results suggest that in certain regimes, increasing structural disorders can induce a phase transition from a trivial insulator to a topological one in the BHZ model. More intriguingly, we find evidence for re-entrant behaviour where the topological phase in the clean limit is destroyed by weak disorder, only to re-emerge again at stronger disorder.

The article is organized as follows. In section \ref{sec:model}, we review a generalization of BHZ model with extended hoppings and describe our amorphization procedure. Section \ref{sec:gap} contains the main results of this paper: we describe the local marker used to characterize the topological nature of the system and present the phase diagrams of the model based on the calculation of the topological invariant.  We further validate our results based on the computation of edge states and conductance in \ref{sec:cond} and summarize our findings in section \ref{sec:conc}.
 \section{The model}
 \label{sec:model}
The BHZ model is a prototypical example of a time-reversal symmetric topological insulator in 2D (class AII) and is commonly used to describe the topological insulating phase observed in HgTe/CdTe quantum wells \cite{bernevig1,Konig}. The model is defined on a square lattice with nearest-neighbor hopping and includes four orbitals per site. Following reference \citep{wchen2}, we adopt the following representation of the gamma matrices:
\begin{equation}
\Gamma^1 = \sigma_z\tau_x, \Gamma^2 = \sigma_0\tau_y ,\Gamma^3 = \sigma_0\tau_z , \Gamma^4 = \sigma_x\tau_x , \Gamma^5 = \sigma_y\tau_x
\end{equation}
Here, $\sigma_i$ and $\tau_i$ are Pauli matrices acting in spin and orbital space, respectively.
In the literature, two main approaches are commonly used to model an amorphous system: (a) a nearest-neighbor-hopping model with a fixed coordination number (for example see \cite{marsal,Cassella2023-nq}) or (b) finite-range hoppings (example see \cite{agarwala1,citian-wang,alvarez,regis}). In the first case, one fixes the coordination number of each lattice site while allowing the exact positions of the sites to vary.  In the second case, hoppings are allowed between any two sites $i$ and $j$ that are within a cut-off distance $R$, typically with a hopping strength that falls exponentially with distance\cite{agarwala1,agarwala2,regis} or Harrison's criteria (power law decay) \cite{cheng, citian-wang}. In this work, we adopt the second approach. The real space Hamiltonian is given by:
\bea{}
H  =  \sum_i \Bigl ( M+4t\Bigl) \dket{r_i}\dbra{r_i} \Gamma^3 + \sum_{i\neq j} \Theta(R- r_{ij})\dket{r_i}\dbra{r_j} T(r_{ij})
\eea
where the summations are over lattice sites. The hopping terms are described by:
\bea{}T(r_{ij}) = f(r_{ij}) \biggl( - \, t \, \Gamma^3 -i\, \lambda \cos{\phi_{ij}} \Gamma^1 -i \, \lambda \sin{\phi_{ij}} \Gamma^2\biggr)
\eea
with $f(r_{ij}) = e^{1 - \frac{r_{ij}}{a}}$ encoding the distance dependence. The parameter $a$ represents the lattice spacing in the crystalline limit and is set to $1$. Thus, $t$ and $\lambda$ represent the nearest-neighbor hopping strength and spin-orbit coupling strength, respectively, for the clean system. When $R=1$, the model reduces to the standard BHZ model defined on a square lattice, where  the bulk gap closes at $M=0$ 
\begin{align}
\label{eq: BHZ_NN}
\nonumber    H &= \sum_i (M+ 4t)\ket{r_i}\bra{r_i}\Gamma^3 - \sum_i t \, \Bigl( \ket{r_i +\hat{e}_x}\bra{r_i} + \\
 \nonumber     & \ket{r_i + \hat{e}_y}\bra{r_i} \, + \, \text{h.c} \,  \Bigl)\, \Gamma^3 \, -  \, \lambda \sum_{i} \Bigl( i\ket{r_i+ \hat{e}_x}\bra{r_i} \, + \, \text{h.c} \, \Bigr) \, \Gamma^1 \\
              &  - \, \lambda \sum_i \Bigl( i\ket{r_i +\hat{e}_y}\bra{r_i} \, + \, \text{h.c} \Bigr) \, \Gamma^2
\end{align}
Here $\hat{e}_x$ and $\hat{e}_y$ denote unit vectors in $x$ and $y$ direction respectively. In experimental setups, the mass parameter $M$ in this effective Hamiltonian—which drives the topological phase transition—can be tuned by adjusting the thickness of the quantum well \cite{bernevig1}. The system is in a topologically non-trivial phase when $M$ is negative, and in a trivial phase when $M$ is positive.

To gradually move away from the crystalline limit, we introduce lattice distortions as follows. We use the standard deviation $\sigma$ of a Gaussian distribution with mean zero as a control parameter. Physically, $\sigma$ quantifies the strength of thermal fluctuations, as the variance is proportional to temperature ($\sigma^2 \propto k_BT$) \cite{citian-wang,franca1}. Each lattice site is independently displaced within a circle of radius $r$, where $r$ is randomly drawn from the Gaussian distribution and the direction is chosen at random. In our study, we vary $\sigma$ from $0$ to $0.40\,a$ in steps of $0.01\,a$. While finite-range hopping models have been used in earlier studies to explore topological phases in amorphous systems, the effect of varying the hopping cutoff distance 
$R$ which controls the spatial extent of connectivity, has received relatively little attention. In this work, we systematically investigate how varying both the structural disorder and the hopping cutoff $R$ affects the topological phase diagram on a square lattice. Notably, we observe that interplay of structural disorder and extent of connectivity can lead to re-entrant phase transitions which to our knowledge, has not been reported previously.

\section{Bulk topological invariant and phase diagram}
\label{sec:gap}
\begin{figure}[!htbp]
  \centering
  \includegraphics[width=0.99\columnwidth,height=6cm]{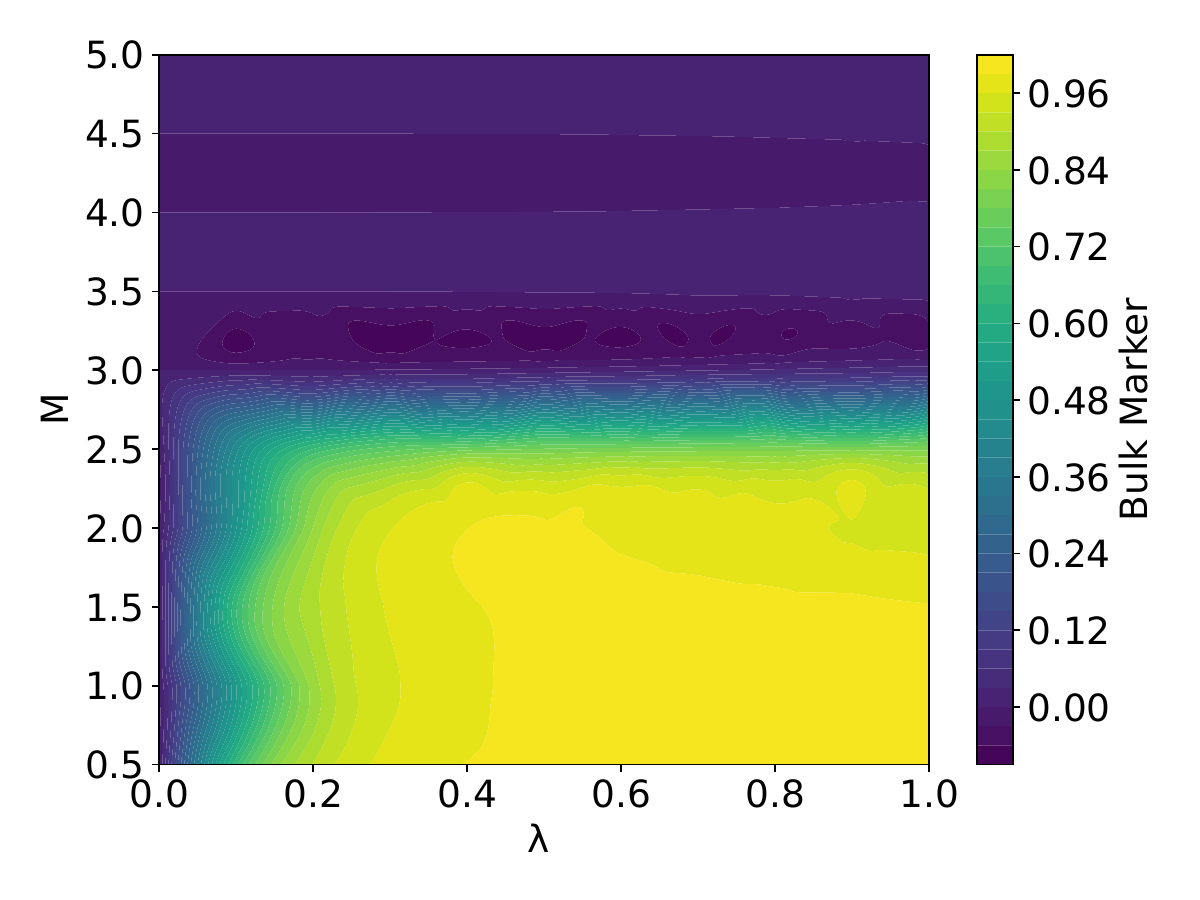}
  \\[4pt] 
  \includegraphics[width=0.99\columnwidth,height=6cm]{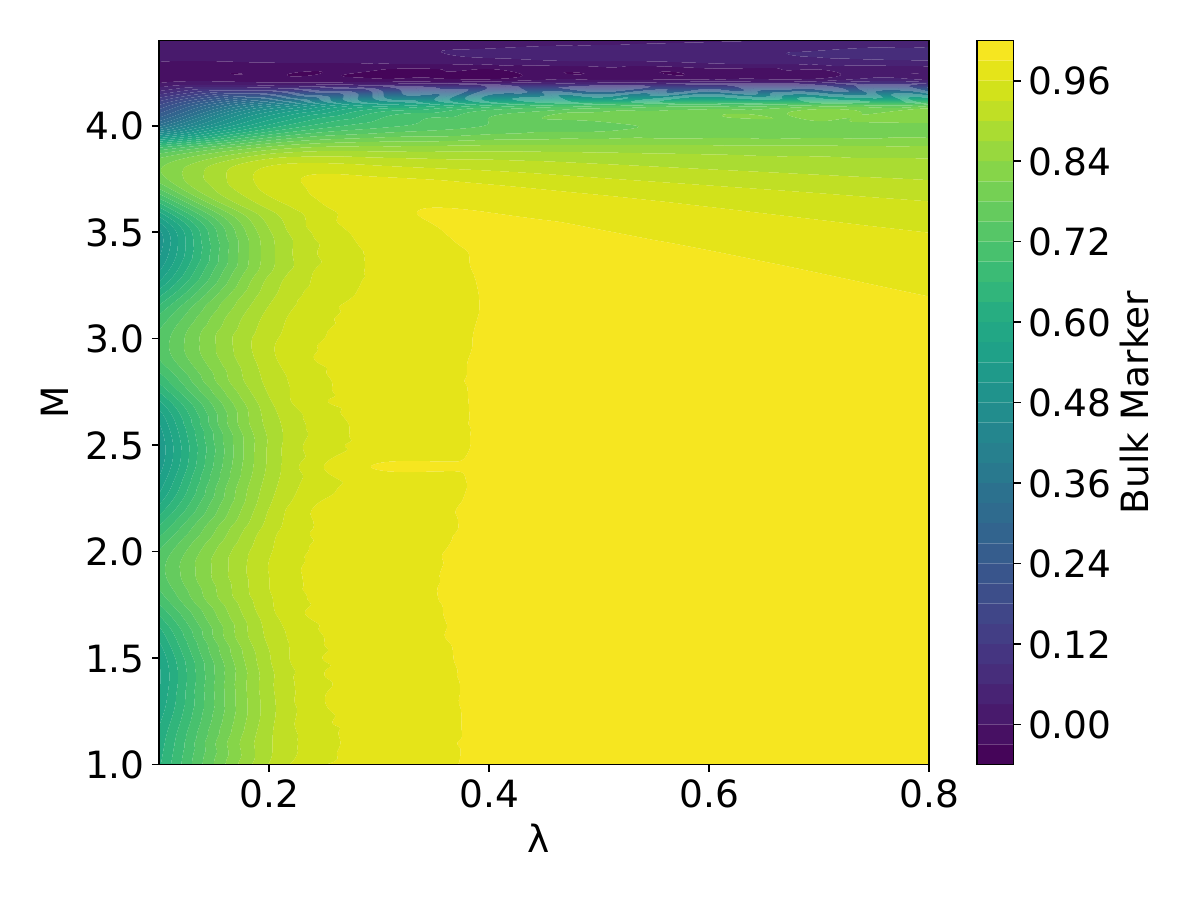}
  \caption{\label{fig:clean}
  Phase diagram for perfect lattice with periodic boundary conditions for
  (a) $R=1.5$, system size: $20 \times 20$, and
  (b) $R=2.0$, system size: $24 \times 24$.
  At $\lambda=0.8$, phase transition occurs at $M \approx 2.62$ for $R=1.5$ and at $M \approx 4.115$ for $R=2.0$.}
\end{figure}
\begin{figure*}[t]
  \centering
{\includegraphics[width=1.0\textwidth,height=0.55\textwidth]{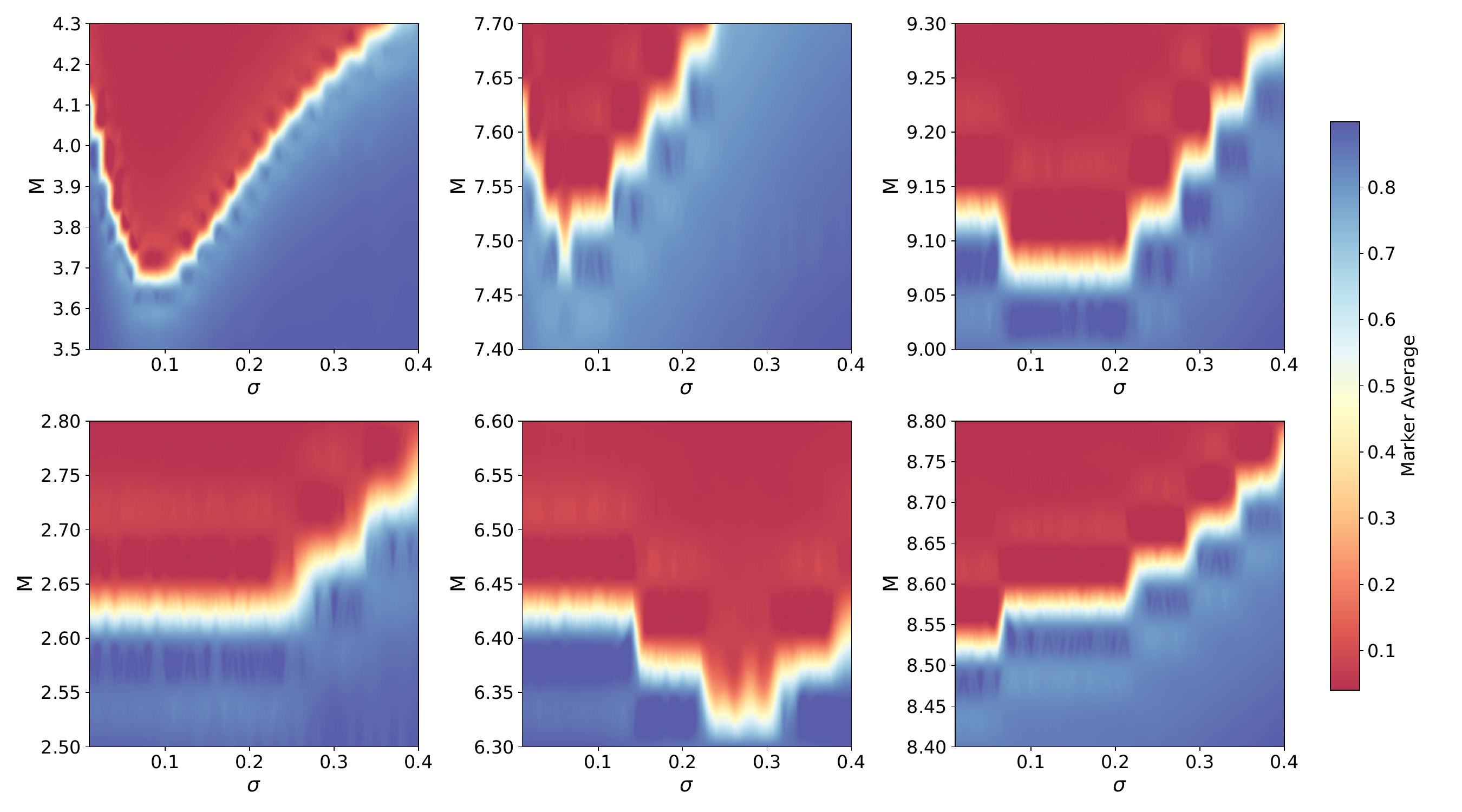}
  }
  \caption{\label{fig:phase_combined}
Phase diagram as a function of $M$ and $\sigma$ at $\lambda=0.8$ with periodic boundary conditions.
  System size is $40 \times 40$.
  Top row: (a) $R=2.03$, (b) $R=3.03$, (c) $R=3.70$.
  Bottom row: (d) $R=1.70$, (e) $R=2.50$, (f) $R=3.38$.}
\end{figure*}
BHZ Hamiltonian is a typical example of what is known as a Dirac Hamiltonian. These are Hamiltonians which have the generic form 
\begin{equation}
\label{eqn:dirac_ham}
    H = \sum_{i=0}^D \, \textbf{v}_i(\textbf{k})\Gamma^i
\end{equation}
in $D$ spatial dimensions. Here $\Gamma^i$ are $N \cross N$ Hermitian matrices satisfying $SO(2n+1)$ Clifford algebra.
\begin{equation}
    \{ \Gamma^i, \Gamma^j\} = 2\delta^{ij}
\end{equation}

For example, the Bloch Hamiltonian corresponding to the real space Hamiltonian in eqn. \ref{eq: BHZ_NN} takes the form 
\begin{equation}
    H(\textbf{k})\,  = \,v_1(k_x) \, \Gamma^1 \, + \, v_2(k_y) \, \Gamma^2 \, + \,  v_3(\textbf{k}) \, \Gamma^3 
\end{equation}
with $v_1(k_x) = 2\lambda \sin{k_x}$, $v_2(k_y)= 2 \lambda \sin{k_y}$ and $v_3(\textbf{k}) = (M+4t) -2t(\cos{k_x} + \cos{k_y})$. When translational symmetry is present, the topological invariants for different symmetry classes can be defined using momentum-space quantities \cite{qi,2963,fu,schnyder}. However, in the presence of disorder, a real-space formulation of the topological invariants is required to diagnose the topological phase \cite{kitaev,bianco,ornellas,Sykes_2022,sykes2,julia,fan}. For Dirac Hamiltonians, it was shown in \cite{gersdorff} that all topological invariants can be expressed in terms of the degree of a map from $T^D$ to $S^D$ in $D$ spatial dimensions. Here, $T^D$, the domain of the map, represents the Brillouin zone in momentum space, while $S^D$, the codomain, corresponds to the Bloch sphere. To see how this map arises, consider the band-flattened Hamiltonian

 \begin{equation}
 \mathbb{Q} = \frac{H}{|\textbf{v}|} = \textbf{n} \cdot \mathbf{\Gamma} \, , \, \textbf{n} = \frac{\textbf{v}}{|\textbf{v}|}
 \end{equation}

Thus, $\textbf{n}$ defines a map from $T^D$ to $S^D$, and the degree of this map is given by
\bea
\deg[\textbf{n}] \, = \, \frac{1}{V_D D!}\int \epsilon_{i_0 \cdots i_D} n^{i_0}\wedge dn^{i_1}\wedge dn^{i_2} \cdots \wedge dn^{i_D}
\eea
In this formulation, for class AII in $2D$, the topological invariant is expressed as:
\bea
\nu = (-1) ^{\deg[\Vec{n}]}
\eea
Subsequently, real-space expressions for topological invariants were derived in \cite{wchen1} and further explored in \cite{oliveira}. In two dimensions, the local marker takes the form:
\bea
\mathcal{C}(\textbf{r}) \, = \, \pi\trace_{\textbf{r}}{W \left(Q\hat{x}P\hat{y}Q - P\hat{x}Q\hat{y}P \right)}
\eea
Here, $W$ is the product of the gamma matrices not appearing in the Hamiltonian, and the trace is taken over all the states localized at site $\textbf{r}$. It should be noted that for systems such as the ones considered in this work where $S_z$ is conserved, this invariant is the same as the spin Chern number \cite{wchen1,wchen2}. In particular, for BHZ model  $W =\Gamma^4\Gamma^5 =  \text{diag}(i,i,-i,-i)$.  $P$ denotes the projector onto single particle eigenstates below $E_F$ and $Q= 1-P$ is the complementary projector. Since the BHZ model with the Fermi level set to $0$ corresponds to a Dirac Hamiltonian, we use this local $\mathbb{Z}_2$ marker to investigate the model’s topological nature.

Before considering the effects of disorder, we first compute the phase diagram of the crystalline system for two representative values of $R$ (Fig. \ref{fig:clean}). For $R = 1.5$, which corresponds to a generalization of the standard nearest-neighbor BHZ model that includes next-nearest-neighbor (NNN) hoppings, the low-energy Hamiltonian of the clean system (for the up-spin block) takes the following form:
\begin{equation}
\label{eq:low-energy-NNN}
H = \begin{pmatrix}
    A +(t+2t')(k_x^2 +k_y^2) & 2(\lambda + \sqrt{2}\lambda')(k_x -ik_y) \\
    2(\lambda + \sqrt{2}\lambda')(k_x +ik_y) &  -A - (t+2t')(k_x^2 +k_y^2)
\end{pmatrix}    
\end{equation}
Here, $A = M - 4t'$, where $t'$ denotes the strength of the next-nearest-neighbor (NNN) hopping, and $\lambda'$ is the corresponding spin-orbit coupling strength. In this case, the system undergoes a phase transition from a topological to a trivial regime at $M = 4t'$. For our choice of parametrization, $t' = te^{1 - \sqrt{2}}$, this gives $M \approx 2.64$. A similar phase transition occurs at $M = 4t' + 4t''$ (i.e., $M \approx 4.115$) for $R = 2.0$, where $t''$ represents the strength of hopping to third nearest neighbors.

\begin{figure}[!htbp]
  \centering
  \includegraphics[width=0.99\columnwidth]{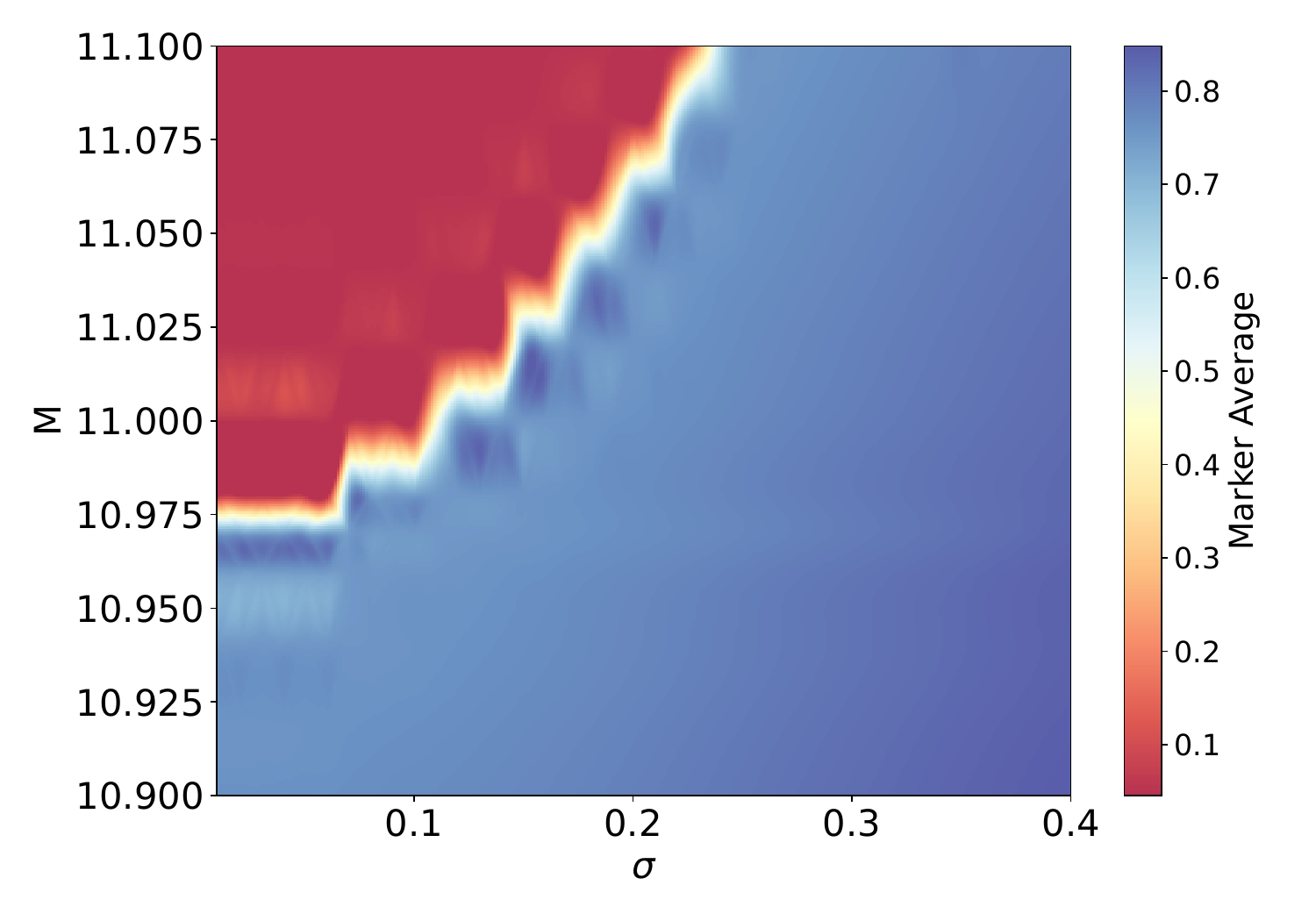}
  \includegraphics[width=0.99\columnwidth]{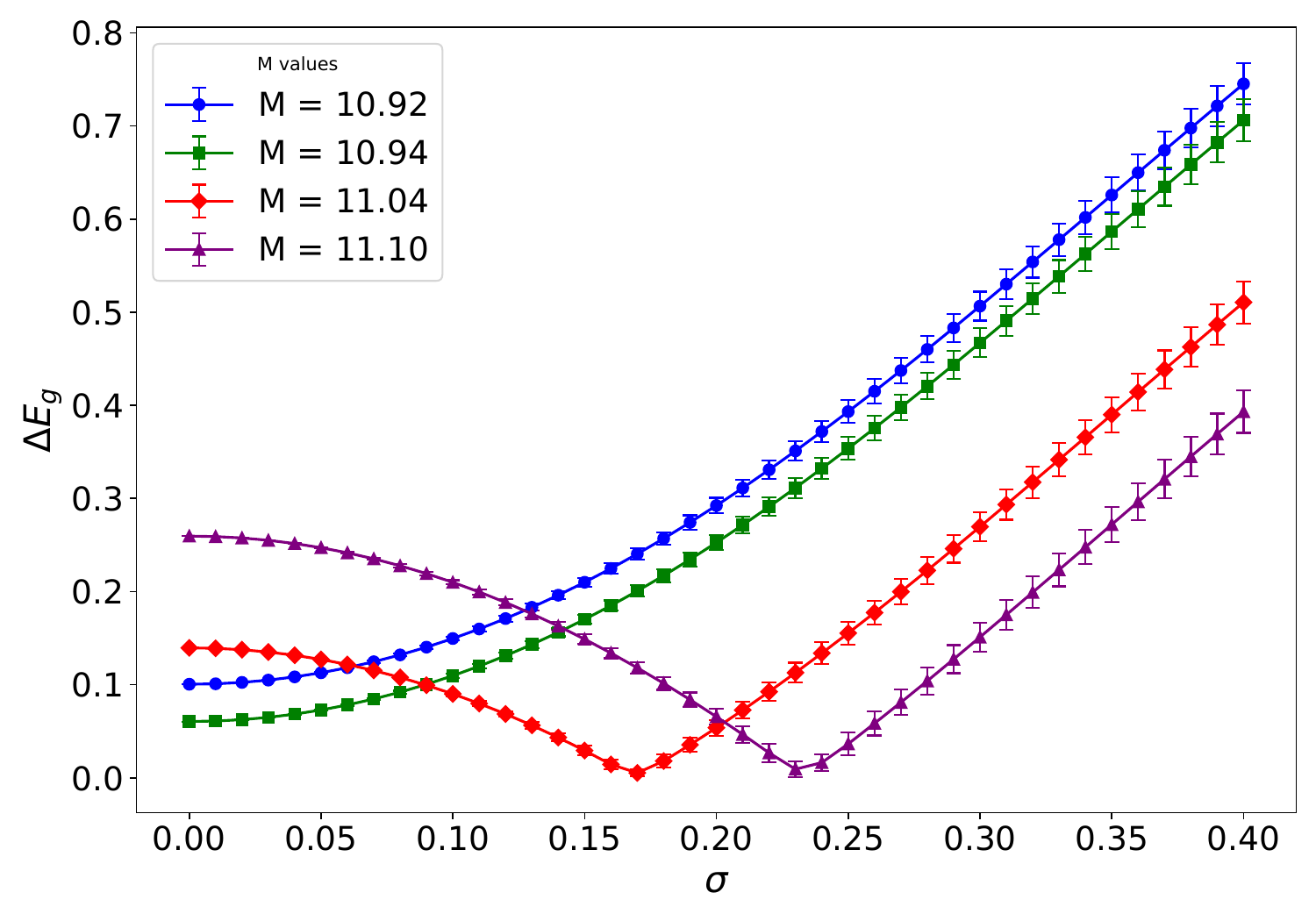}
  \caption{\label{fig:R24.03}
 Top panel: Topological phase diagram for $R=24.03$ ($\approx$ no cutoff limit) Bottom panel: Average bulk gap as a function of $\sigma$ for $M=10.92$ (blue), 10.94 (green), 11.04 (red) and 11.10 (purple).}
\end{figure}
For the usual Anderson-type disorder, it is known \cite{Li1} that near the phase boundary, disorder can induce a transition to a topologically non-trivial state in the BHZ model. In contrast, reference \citep{song1} studied the effect of random bond strengths in the same model on a perfect lattice and found that such disorder does not induce a transition from a trivial to a non-trivial phase. To examine whether this conclusion remains valid for a model of an amorphous system with extended hoppings, we focus on parameter values near the phase boundary of the corresponding crystalline system. For this analysis, we consider a $40 \times 40$ system and evaluate the bulk topological marker by averaging over 20 disorder configurations. Since the marker converges to its quantized value more rapidly for larger spin-orbit coupling, we study the effect of amorphization at $\lambda = 0.8$. 

It is important to note that values of $R$ that coincide with the exact distance to certain $n$th-nearest neighbors in the clean lattice should be avoided (see appendix-\ref{sec:app-C} for a more detailed discussion on this point \cite{alvarez_2022}). Here we present results for two categories of $R$ values. In the first category, $R=2.03$, $3.03$ and $3.70$ are relatively close to the distances of specific $n$th-neighbors in the square lattice, which are $2$, $3$ and $\sqrt{13}$ respectively. In the second category, $R=1.70$, $2.50$ and $3.38$ lie approximately $0.3$ units away from the hopping bond distances closest to $R$. For example, the two bond distances closest to $R=1.70$ are $\sqrt2$ for 2nd nearest neighbors and $2$ for 3rd nearest neighbors. We note that this distinction is meaningful primarily for small values of $R$, because as $R$ increases, the difference between successive nearest-neighbor distances ($d_n$ -$d_{n-1}$) decreases. As a result, the notion of avoiding specific $n$th-neighbor distances become less well-defined. However, this is not problematic at large $R$, since the hopping amplitudes decay exponentially with distance, and contributions from sites near the cut-off are already negligibly small.

\begin{figure*}[t]
  \centering
{\includegraphics[width=0.99\textwidth,height=0.30\textwidth]{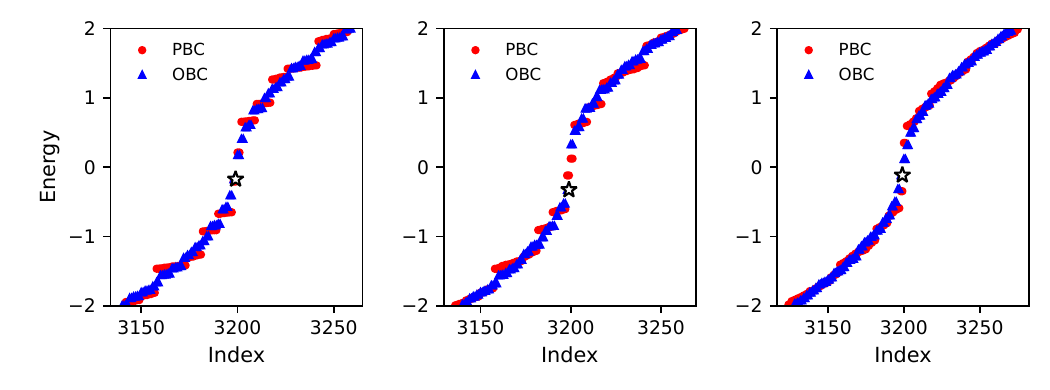}
  }\\

  \subfigure[]{\includegraphics[width=0.99\textwidth,height=0.30\textwidth]{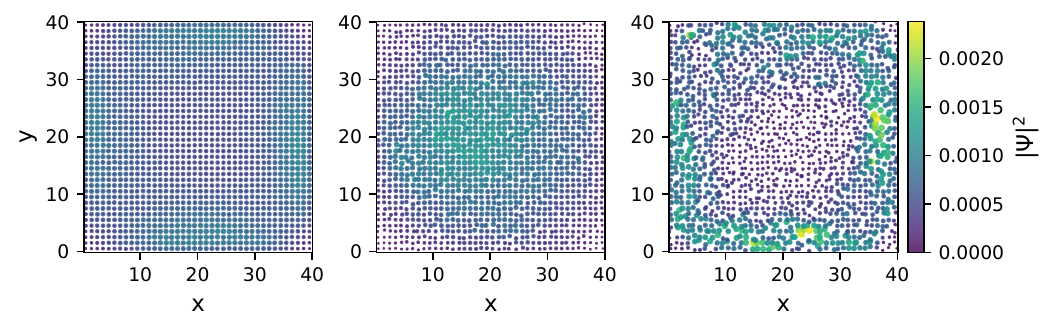}
  }

  \caption{\label{fig:edge_R2}
  Top panel: Spectrum for $R=2.03$ with periodic and open boundary conditions at different values of $\sigma$ for $M=3.80$ close to $E=0$ for a single configuration.
  Red dots denote the states for periodic boundary conditions, while blue dots denote states for open boundary conditions.
  Bottom panel: Plot of $|\Psi|^2$ for the state marked by a star in the figure to its right.
  (a,d) $\sigma=0.02$, (b,e) $\sigma=0.10$, (c,f) $\sigma=0.30$.
  The colour and size of the blobs are proportional to the wavefunction density at each site.}
\end{figure*}
For $R=2.03$, which is close to the third  and fourth nearest neighbor distances (2 and $\sqrt{5} \approx 2.24$) the clean system undergoes a topological-trivial phase transition near $M$ $\approx$ 4.11 (see Fig. \ref{fig:phase_combined} a). The topological phase remains stable for small disorder strength but is driven into a trivial phase as disorder increases. Upon further increase in disorder strength, the system undergoes another transition into a non-trivial insulating phase. This re-entrant nature of phase transition can be also seen from the conductance, which we present in section \ref{sec:cond}. Similar re-entrant behaviour is observed for $R=3.03$ and $R=3.70$, although the precise value of disorder strength ($\sigma$) at which the phase transition occurs depends on the value of the mass parameter $M$. Furthermore, when the clean system lies deep in the topological phase, it remains stable upto $\sigma =0.40$ for each value of $R$. 

We now consider the case where $R$ is not close to any $n$th-neighbor distance of the square lattice, focusing on the phase diagrams for $R=1.70$, $2.50$ and 
$3.38$, shown in the lower panel of Fig.~\ref{fig:phase_combined}. An important aspect in which $R=2.50$ differs from the other two cases is that close to the phase boundary, a topological system can be driven to a trivial phase by structural disorder. However compared to $R=2.03, 3.03$ and 3.70, the topological phase is stable up to larger disorder strength ($\sigma \approx 0.15$). Interestingly, re-entrant phase transition is also observed only at $R=2.50$. In this case however, the structural disorder is not sufficient to drive the trivial phase into a topological one. 

Finally, we consider the case where no hard cut-off is imposed and the distance dependence of hopping is fully captured by an exponential decay. In this context, we chose $R=24.03$ as a practical representation of the no cut-off limit and the corresponding phase diagram is presented in fig. \ref{fig:R24.03}. Since $e^{1-18} \approx 10^{-8}$, the hopping amplitudes beyond this range are negligible - on the order of machine precision for complex floating point calculations. Thus, $R=24.03$ provides a numerically accurate approximation to $R \rightarrow \infty$ limit. We notice that when the clean system is in topological phase ($M < 10. 97$), it remains stable for the entire range of disorder strength considered in this study ($\sigma \leq 0.40$). Conversely, if the clean system lies in the trivial phase, disorder can drive into a topological regime via amorphization. As observed for smaller values of $R$, the further the system is from the phase boundary (i.e., the larger the mass parameter), the stronger the disorder required to induce such a transition .

Although, our study is based on the model of a 2D topological insulator, we note that the disappearance of the topological phase at small disorder and its reappearance at large disorder as seen in $R=2.03, 2.5$ and $3.03$ is reminiscent of the behaviour seen in $\text{Bi}_2\text{Se}_3$ films in \cite{Corbae2023-kq}, where the topological nature of the amorphous samples was investigated by a combination of ARPES/SARPES and transport experiments. In \cite{Corbae2023-kq}, the authors reported that while the perfect crystal was a topological insulator, the nanocrystalline sample showed insulating resistivity and a featureless ARPES. In contrast, the amorphous sample demonstrated spin-polarized surface states. To theoretically investigate the existence of topological phase in the absence of crystalline symmetry, \cite{Corbae2023-kq} studied a model where the sites are distributed randomly keeping the density fixed and rejecting sites which are closer than one lattice constant. The Hamiltonian was based on the regularized version of the low energy Hamiltonian of 3D $\text{Bi}_2\text{Se}_3$ with only nearest neighbor hoppings  (hence a coordination number of 6), with randomness arising from the phase of the spin-orbit coupling term only (the hopping strengths were kept fixed). Using this model, it was shown that the non-crystalline system can exhibit both trivial and non-trivial topological phase depending on the value of the mass parameter. However, such a model does not allow one to study the influence of gradually increasing the structural disorder in the system - a question that can be systematically explored within the framework adopted in this work.
\begin{figure*}[t]
  \centering
{\includegraphics[width=1.0\textwidth,height=0.55\textwidth]{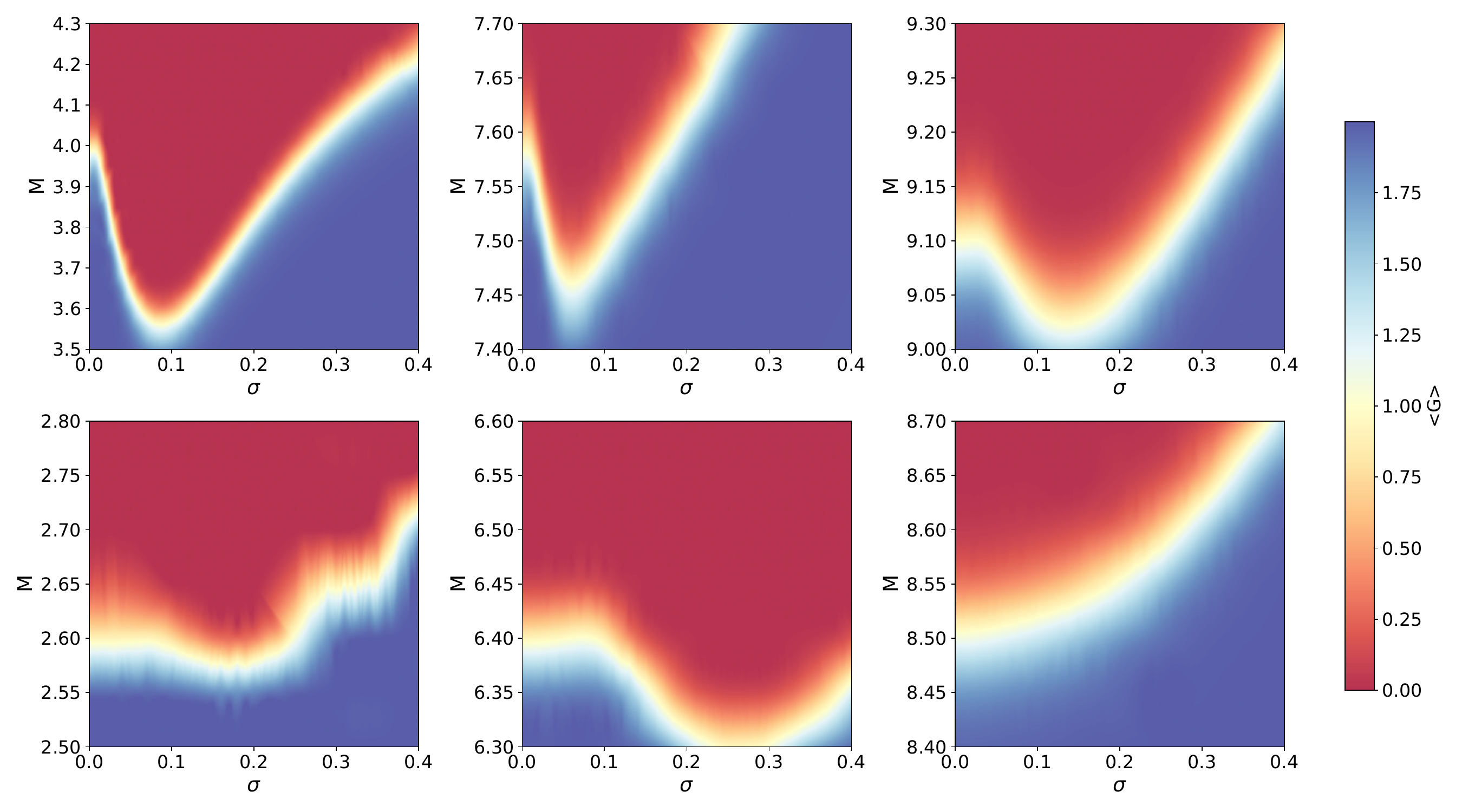}
  }
  \caption{\label{fig:pd_cond}
  Phase diagram for the amorphous BHZ model based on two-terminal conductance.
  Top row: (a) $R=2.03$, (b) $R=3.03$, (c) $R=3.70$.
  Bottom row: (d) $R=1.70$, (e) $R=2.50$, (f) $R=3.38$.}
\end{figure*}

\section{Edge states and conductance}
\label{sec:cond}

A hallmark of a topologically non-trivial state is the bulk-boundary correspondence: if the bulk is topologically non-trivial, there must exist gapless edge states at the boundary. In this section, we validate the results from the previous section by exploring the boundary physics of the system. In figure \ref{fig:edge_R2} we present the energy spectrum and site-resolved wavefunction density for $R=2.03$ ($M=3.80$). From the calculation of the topological marker, we expect that system is in a non-trivial phase at $\sigma=0.02 $ and $\sigma=0.30$, whereas it is in a trivial phase for $\sigma=0.10$. We note that while mid-gap states appear with open boundary conditions at $\sigma=0.02$ and $0.30$, they are not truly gapless due to the finite size of the system. However, the distinct nature of the states is clear once we look at the density distributions of the wavefunction. For both $\sigma=0.02$ and $\sigma=0.30$, the mid-gap states are localized on the boundary of the sample. In the intermediate disorder regime ($\sigma=0.10$), since the gap with periodic boundary conditions is smaller than the one with open boundary conditions, we plot $|\psi|^2$ for a state with negative energy with the least absolute value. Not surprisingly, it is spread throughout the center of the sample confirming that the system is in a trivial phase.
To validate our findings, we compute the two-terminal conductance using Landauer-Buttiker formalism\cite{landauer,buttiker} which allows us to link the system's topological properties with experimentally measurable transport properties. By examining the variation of conductance with disorder, one can directly observe how disorder affects the topological phase and hence it provides  a more direct connection between theory and experiment. In the Landauer-Buttiker approach, the conductance is obtained from the transmission matrix $T$ of the system which encodes the probability of electrons being transmitted through the system. More precisely, the conductance is given by:
\bea
G(E) = \frac{e^2}{h}\trace(TT^{\dagger})
\eea
For these calculations, we have used the KWANT package\cite{groth1}. The leads are modeled as translational invariant systems with $4$ orbitals per site and nearest-neighbor hopping. The onsite energies and hopping amplitudes in the leads are chosen sufficiently large to ensure metallic behavior at $E_F=0$, and the hopping terms involve only the $\Gamma^3$ matrix of the BHZ Hamiltonian. For conductance, one needs to go to much larger system sizes in order to obtain the phase diagram reliably. In particular, we have observed that this depends on the extent of the hoppings of the system (value of hard cut-off). Fig. \ref{fig:pd_cond} shows the phase diagram on the basis of conductance for the same values of $R$ which we have considered previously, where the system size ranges from a $144 \cross 144$ sample for $R=2.03$ to $264 \cross 264$ sample for $R=3.7$. It can be seen that the phase diagrams for both $R=2.03$ and $R=3.03$ are nearly identical to the ones obtained by computing the bulk invariant ((a) and (b) in \ref{fig:phase_combined}). There is a small discrepancy near the phase boundaries. Focusing our attention on $R=2.03$, we note that for $M=4.15$ and $M=4.20$, a phase transition to a non-trivial state at large disorder is expected, but this does not seem to be true from the phase diagram obtained on the basis of conductance. We attribute these discrepancies to the small size of the system. This is evident from the observation that even at $M=4.10$ - a parameter point which is known to lie in the topological regime in the clean limit - the conductance is close to $0$. This suggests that the system size is insufficient to capture the behaviour of edge transport characteristic of the topological phase, which is expected as the correlation length diverges as one approaches the phase transition point. For other values of $R$, even though the phase diagrams do not match exactly, they still retain the characteristic features expected from the marker calculation. In particular, they confirm that the re-entrant phase transitions will be seen for precisely those values of $R$ which were expected to show such behaviour on the basis of local marker. Furthermore, we also find that for $R=1.70$ and 3.38, a system which was initially topological continues to remain so as the disorder strength is increased.
\section{Off-block diagonal terms}
\begin{figure}[!htbp]
  \centering
  \includegraphics[width=0.95\columnwidth,height=6.5cm]{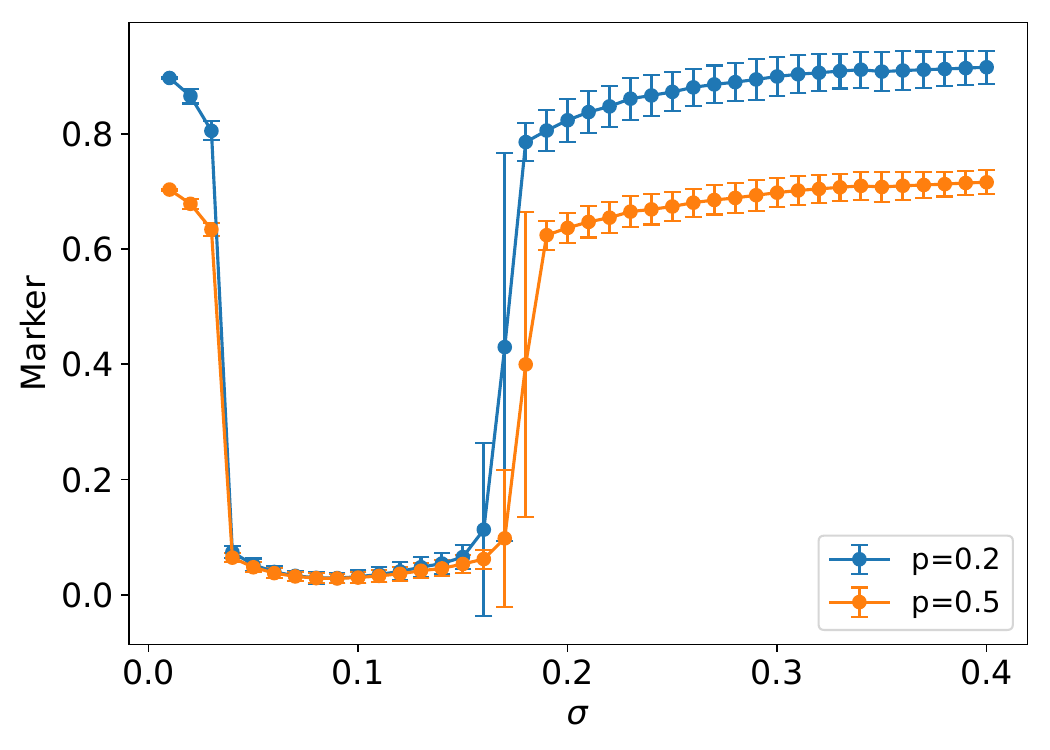}
  \includegraphics[width=0.95\columnwidth,height=6.5cm]{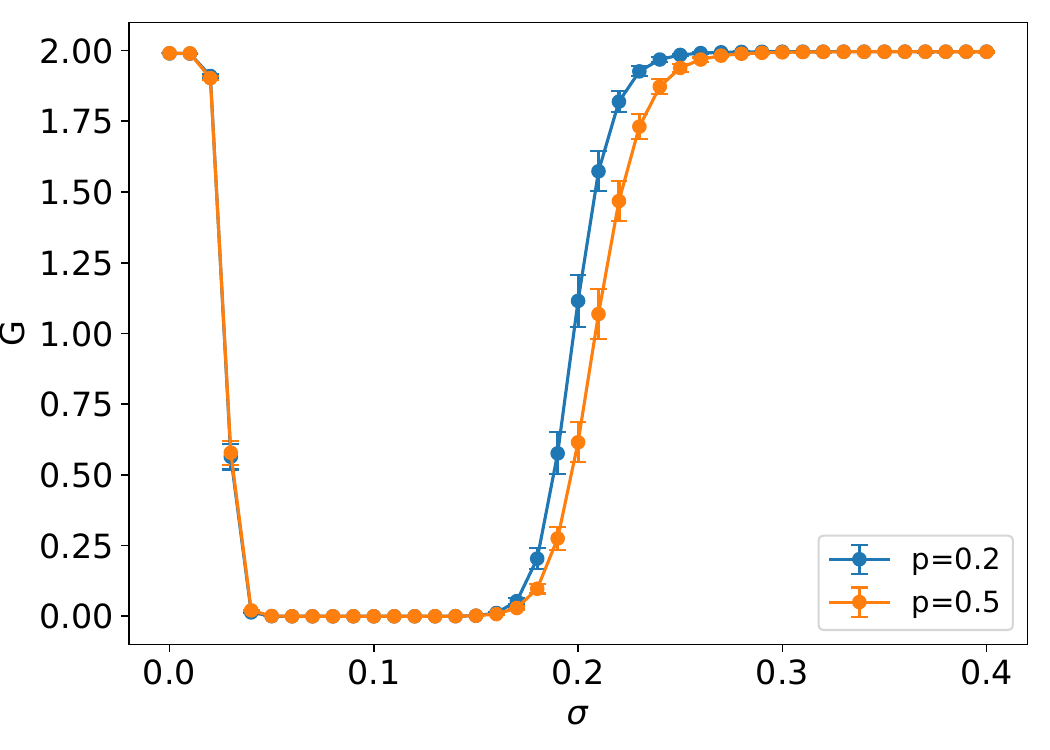}
  \caption{\label{fig:perturbation_coupled}
 Plot of mean marker (top) vs $\sigma$ and  mean conductance vs $\sigma$ (bottom) for $H'$ with $R=2.03$ and $M=3.85$. The marker value decreases as the strength of the $\Gamma^4$ (p) increases for the same value of disorder strength ($\sigma$) while mean conductance remains quantized.}
\end{figure}

We now discuss the impact of introducing a term which couples the up and down spin so that $S_z$ is no longer conserved. For example, one can include a term proportional to $\Gamma^4=\sigma_x\tau_x$ and is an odd function of $\textbf{k}$, to the usual nearest-neighbor BHZ model
\begin{align}
\label{eqn:H_prime}
\nonumber H' \, &= \, H_{\text{BHZ}} - 2p\,(\sin{k_x} +\sin{k_y}) \, \Gamma^4 \\
                &=  v_1(k_x)\Gamma^1 + v_2(k_y)\Gamma^2 + v_3(\textbf{k})\Gamma^3 + v_4(\textbf{k})\Gamma^4  
\end{align}
This Hamiltonian remains in the same symmetry class while breaking the $U(1)$ symmetry associated with $S_z$ conservation. However, one can see that the Hamiltonian $H'$ is not a Dirac Hamiltonian of the form given in eqn. \ref{eqn:dirac_ham} because in $D=2$, such a Hamiltonian can only consist of three gamma matrices. As a result, the local $Z_2$ marker is not expected to be a true topological invariant for this system. In order to check the behaviour of the marker in the presence of the additional term, we compute the marker for $R=2.03$ and $M=3.85$ for two different values of $p$ (see Fig. \ref{fig:perturbation_coupled}, top panel). It can be seen that the marker can still distinguish between the two phases for relatively small values of $p$: it shows a finite (non-integer) value in the topological phase which decreases as the strength of the coupling $p$ is increased. This is confirmed by the two-terminal conductance calculated for the same model with the same parameters which shows a quantized conductance of $2e^2/h$ in the regions where the marker shows a finite value, indicating that the phase-diagrams calculated for the amorphous BHZ models are expected to be valid even in the absence of $S_z$ conservation. Thus, the decrease in the value of $Z_2$ marker in this model is a reflection of the fact that $H'$ is not a Dirac Hamiltonian and not of the fact that $S_z$ is no longer conserved. We further verify this, by considering the following Hamiltonian closely related to the BHZ model 
\begin{equation}
H_{\text{BHZ-like}} = v_1(k_x)\Gamma^1 +v_3(\textbf{k})\Gamma^3 + v_4(\textbf{k})\Gamma^4
\end{equation}
where $v_1$, $v_3$ and $v_4$ are defined as in eqn. \ref{eqn:H_prime}. Just like the BHZ Hamiltonian, this has just three gamma matrices but the presence of $\Gamma^4$ ensures that there are off-block diagonal terms. At the same time, the gap closing point is not shifted which makes it easier to compare with the results of the BHZ model. Hence, it provides an ideal test-ground for the most general use of the local $Z_2$ marker. It should be noted that the operator $W$ in the local marker is given by product of $\Gamma^2$ and $\Gamma^5$ for this Hamiltonian. We obtain the phase diagram for $R=2.03$ and $p=0.8$  and find that it is nearly identical to the one obtained for the generalized BHZ model. 

\begin{figure}[!htbp]
\includegraphics[width=0.99\columnwidth, height=6cm]{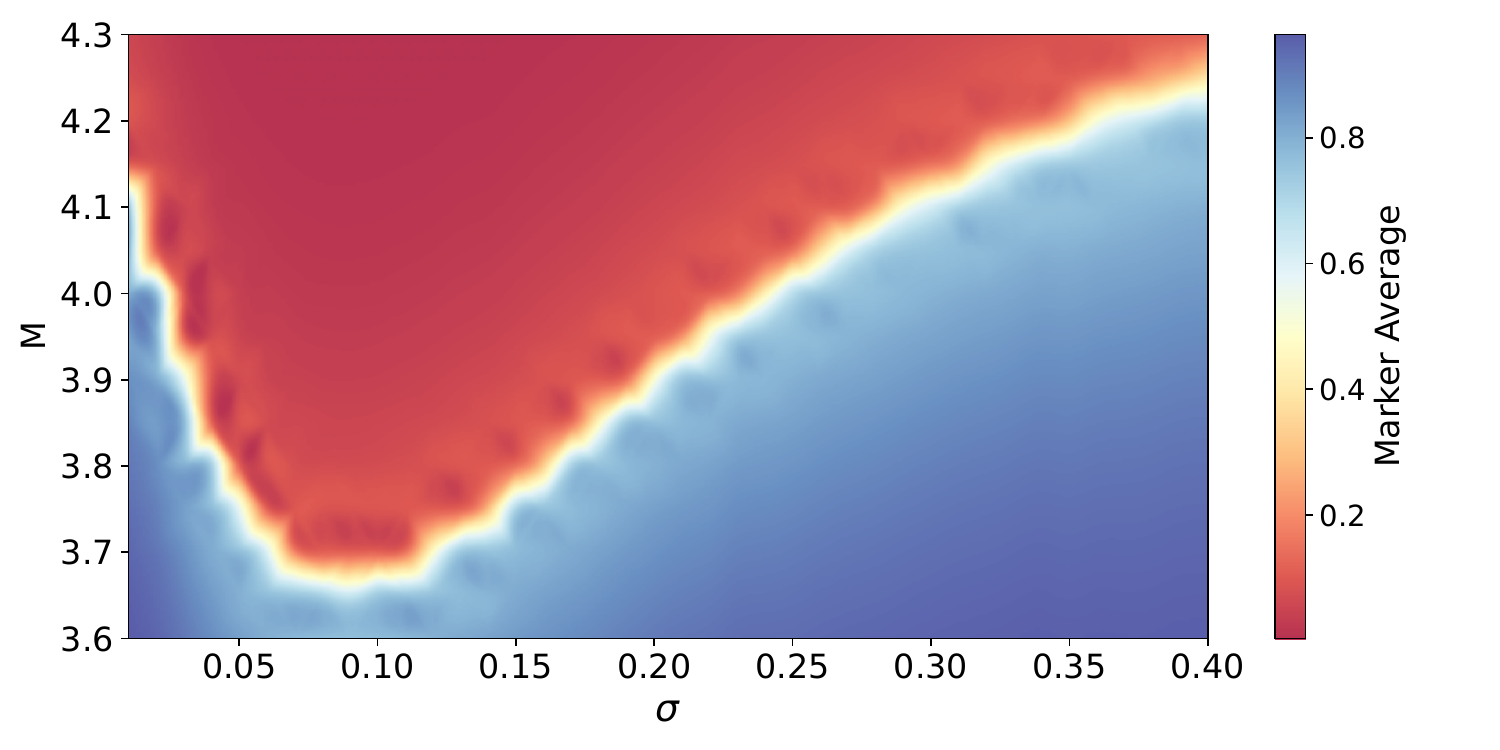}
\caption{\label{fig:R2.03_new_ham}Phase diagram based on local marker for $R=2.03$ for $H_{\text{BHZ-like}}$}
\end{figure}

\section{Conclusion and outlook}
\label{sec:conc}
In this work, we have studied the phase diagram of the amorphous BHZ model, focusing on parameter regimes close to the phase transition point in the clean crystalline system. We have found that different short-ranged models, as parameterized by the extent of hopping in the clean system, can exhibit different topological phase diagram when tuning the strength of the structural disorder. In particular, there exist parameter regimes where the system can exhibit re-entrant phase transition. The model exhibits a general trend where increasing the hoppings tends to stabilize the topological phase and in the limit of no cut-off, the topological phase remains stable for the entire range of the disorder strength considered. Looking forward, we envisage two specific and natural questions which deserve further investigation. Firstly, the role of extended hoppings in amorphization of topological insulators described by microscopic Hamiltonians and defined on different lattices needs to be elucidated. Secondly, it would be interesting to study the critical properties of the system near the two topological phase transitions in systems exhibiting re-entrant phase transition and to contrast them with Anderson-disorder driven phase transitions. We hope to address these in future work.

\acknowledgments{This work is supported by Center for Emergent Materials at the Ohio State University, a National Science Foundation (NSF) MRSEC through NSF Award No. NSF DMR-2011876.}

\section*{Data Availability}
The data that support the findings of this study were generated by numerical simulations. The source code and parameters used to generate the simulations is publicly available. \cite{data}.

\appendix

\section{Disorder average}

\begin{figure}[!htbp]
\includegraphics[width=0.99\columnwidth,height=6cm]{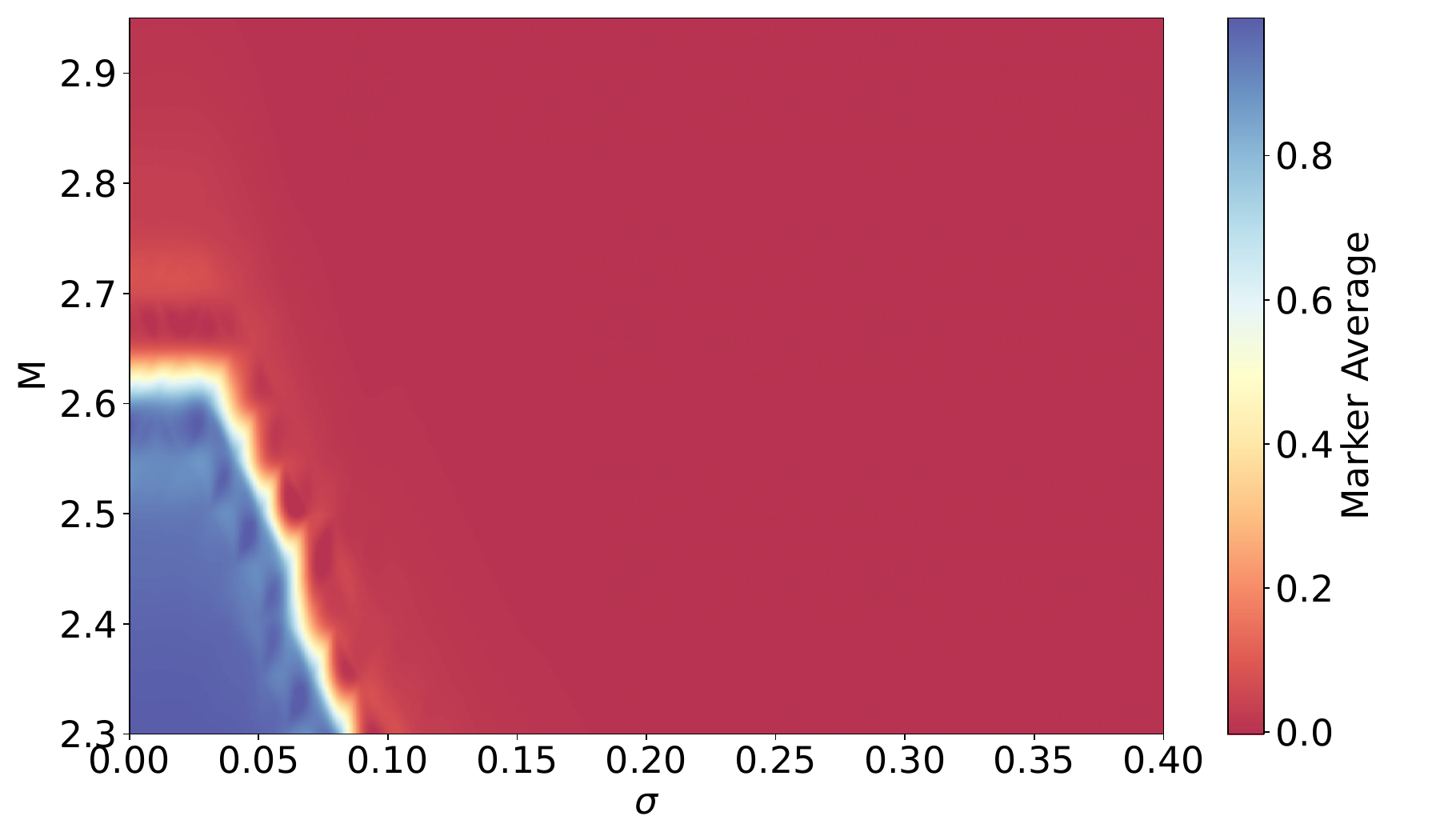}
\caption{\label{fig:R1.5}Phase diagram based on local marker for $R=1.50$}
\end{figure}

\begin{figure}[!htbp]
  \centering
  \includegraphics[width=0.95\columnwidth,height=6.cm]{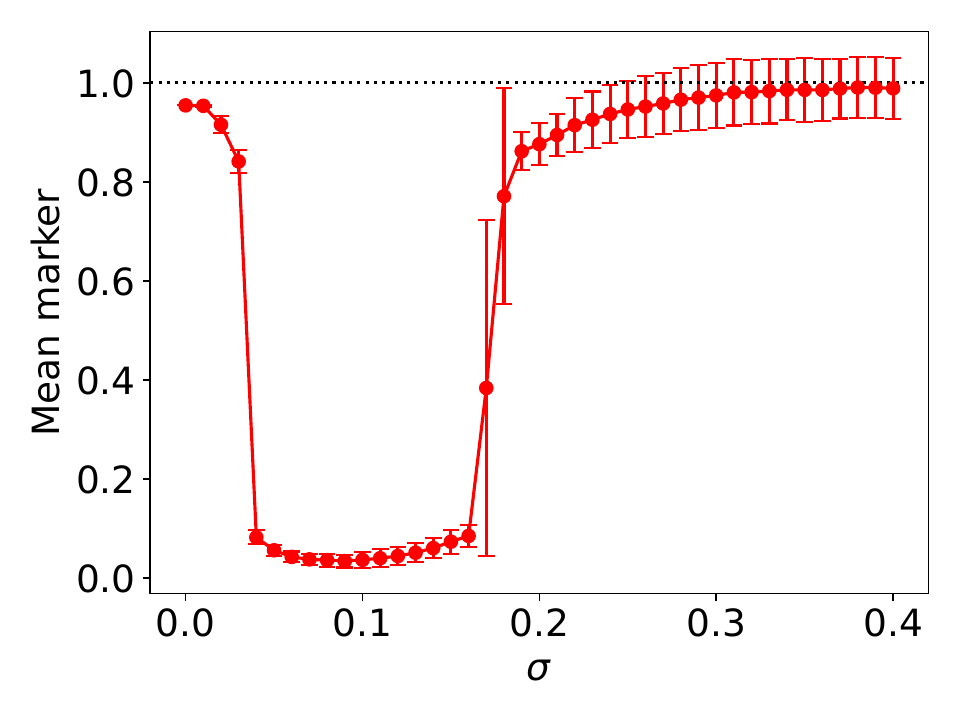}
  \includegraphics[width=0.95\columnwidth,height=6.cm]{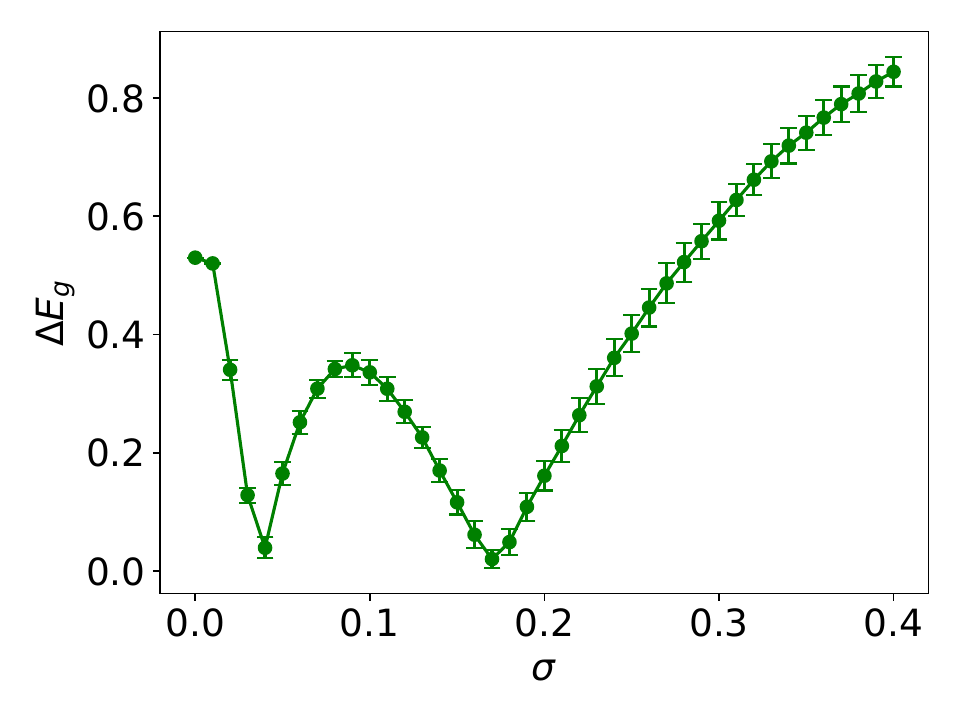}
  \caption{\label{fig:R2.03gap}
 Top : Mean marker. Bottom : Spectral gap with periodic boundary conditions. $R=2.03$ and $M=3.85$}
\end{figure}

To show the magnitude of fluctuations after disorder averaging, we present the bulk spectral gap and the mean marker along a horizontal cut of the phase diagram for $R=2.03$. For this purpose, we chose a value of $M$ (3.85) which exhibits re-entrant transition as the disorder strength is varied. Fig. \ref{fig:R2.03gap} shows the mean marker and bulk gap averaged over 20 configurations as ta function of the strength of the disorder ($\sigma$). The magnitude of fluctuations of the bulk gap remain of the same order for all values of $\sigma$. However, this is not the case for the mean marker, which shows large fluctuations in the transition region (between $0.15 \lesssim \sigma \lesssim 0.20$ ). It is also interesting to observe that since the average bulk gap is larger at $\sigma=0.40$ in comparison to $\sigma=0.00$, the mean value of marker is also higher in the disordered topological phase ($\approx 0.99$) compared to the clean topological insulator (0.95) . 

\section{Phase diagrams for additional values of R}    
In the main text, barring the case of $R=2.50$ ($M=6.40$), one either observes that the topological phase persists throughout the full range of disorder strength or undergoes a re-entrant phase transition. However, one can also encounter a situation where the system neither exhibits re-entrant phase transition nor is it stable at large structural disorder. This can be seen in figure \ref{fig:R1.5}, where we present the phase diagram for $R=1.5$. There are two main features in this case. First, as $M$ is decreased from $2.64$ (the phase transition point), the system remains in the topological phase even at higher disorder strengths. Second, for $M> 2.64$, the system remains trivial for all values of $\sigma$. Therefore, for $R= 1.5$, structural disorder does not induce a transition from the trivial to the topological regime, similar to the case of nearest-neighbor BHZ model with bond disorder \cite{song1}.

\begin{figure}[!htbp]
  \centering
  \includegraphics[width=0.99\columnwidth,height=6cm]{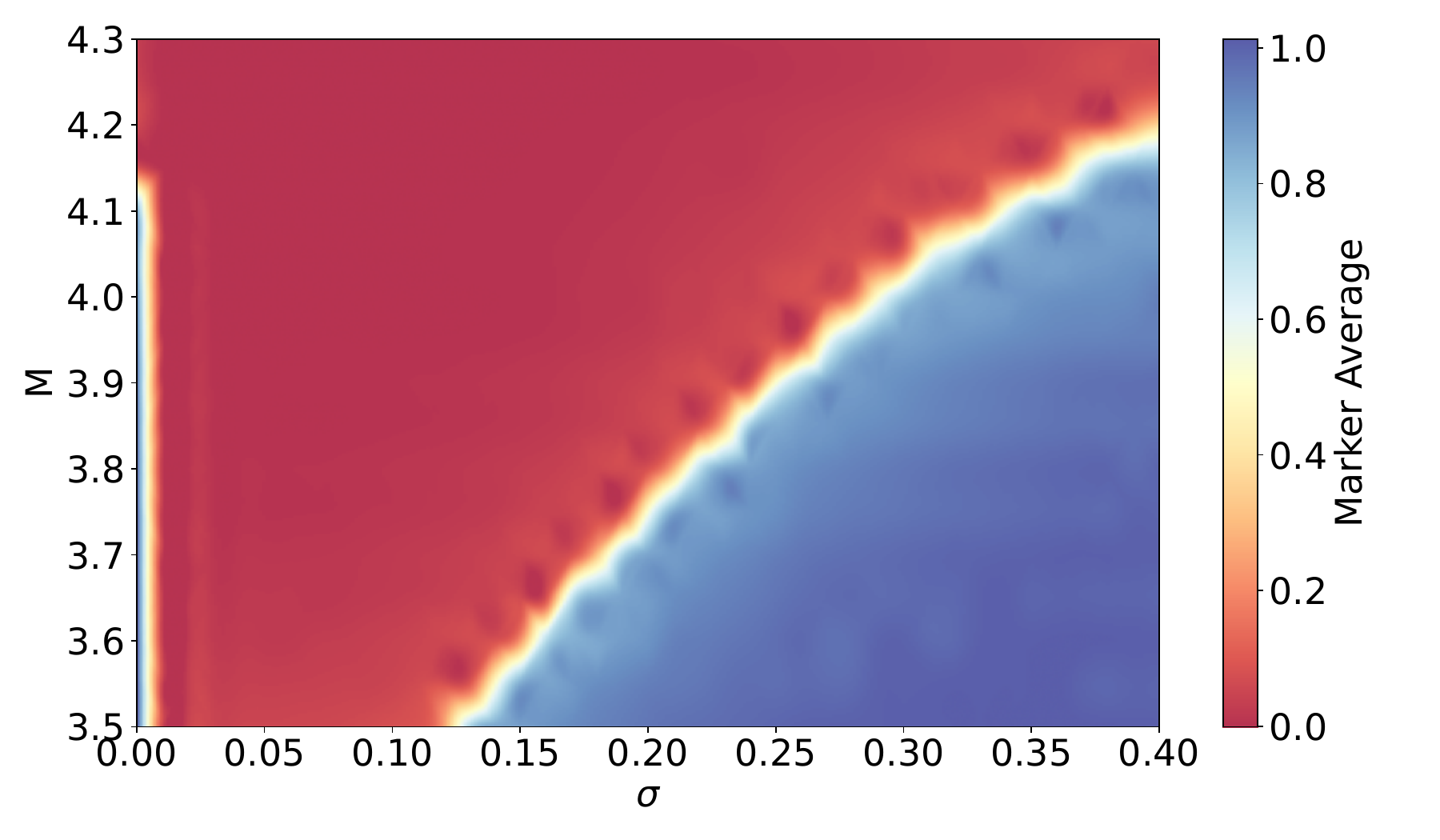}
  \includegraphics[width=0.99\columnwidth,height=6cm]{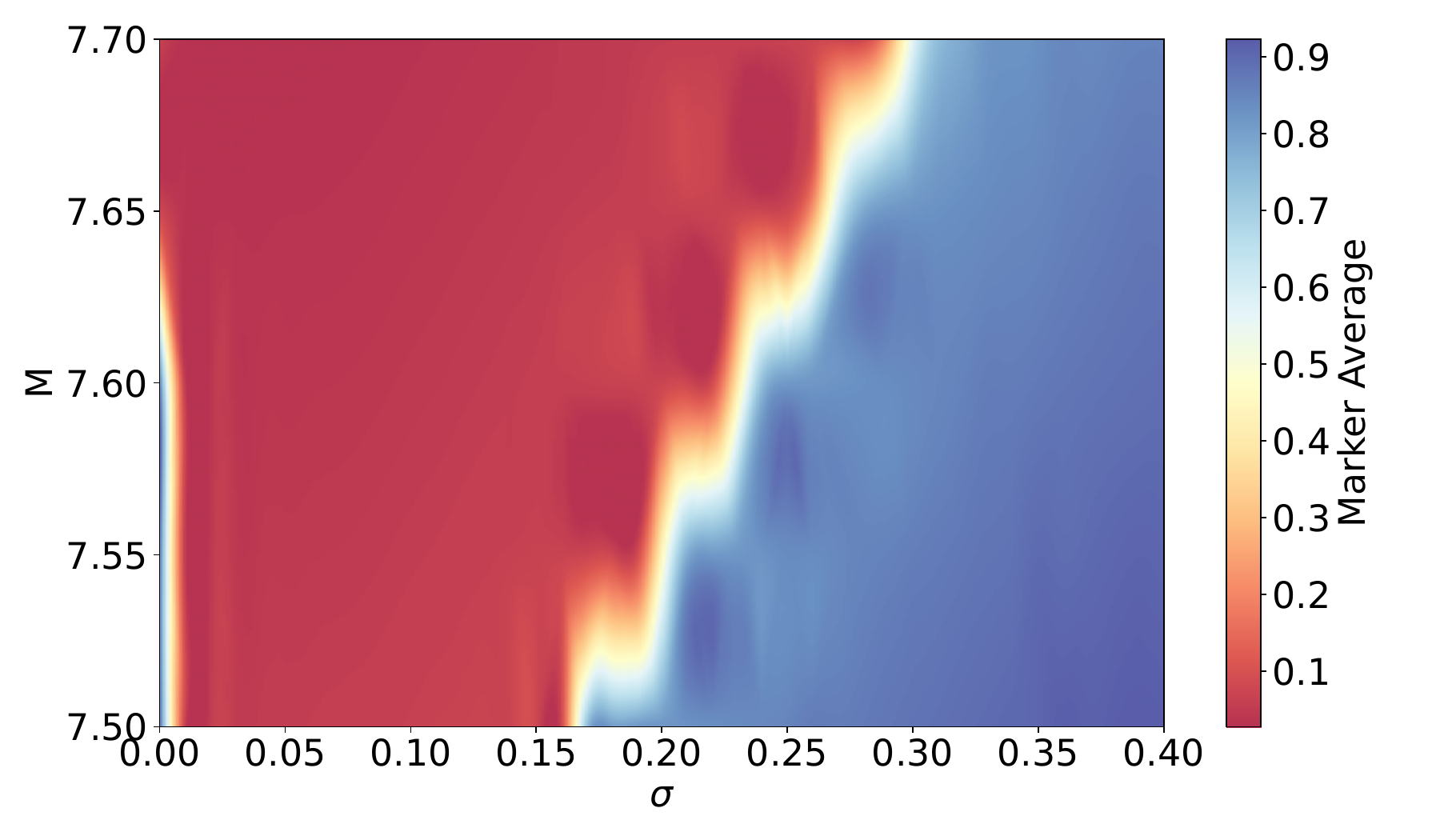}
  \caption{\label{fig:cut_off}
Phase diagrams based on local marker. Top : $R=2.00$, Bottom : $R=3.00$}
\end{figure}
\section{Restriction on the choice of hard cut-off } 

\label{sec:app-C}

\begin{figure}[!htbp]
\includegraphics[width=0.99\columnwidth]{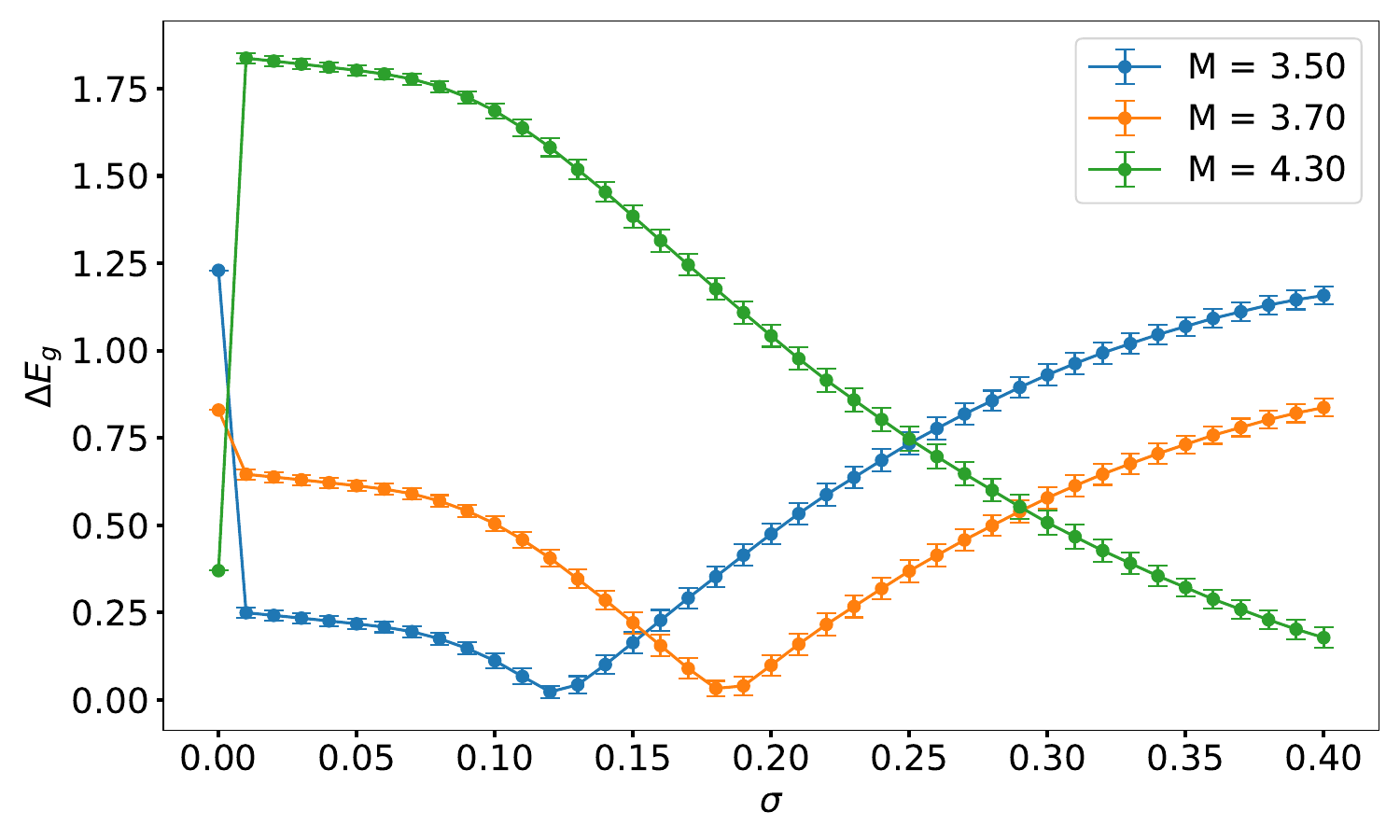}
\caption{\label{fig:R2gap} Spectral gap with periodic boundary conditions for $R=2.0$ at $\lambda=0.8$. }
\end{figure}

In this section, we comment on some technical aspects regarding the choice of hard cut-off $R$ in our model. In the main text, we mentioned that one should avoid choosing values of $R$ which coincides with some $n^{th}$ nearest neighbor distance in the perfect lattice. We illustrate this issue by presenting the phase diagram for $R=2.0$ and $R=3.0$ here. The topological phase exists along a narrow line near $\sigma = 0.0$. Even a small amount of disorder drives the system into a trivial phase, but the topological phase re-emerges as disorder is further increased. The presence of a topological phase in the crystalline limit and its destruction by an arbitrarily small amount of disorder seems paradoxical. To further examine this phenomenon, we computed the bulk spectral gap  as a function of disorder strength (see Fig.~\ref{fig:R2gap}). Interestingly, we observe a sharp jump in the bulk gap between $\sigma = 0.00$ and $\sigma = 0.01$, regardless of the value of $M$. We have checked that this behaviour persists even when we set $\sigma=10^{-6}$. This strongly suggests that even though the disorder is small, the change in Hamiltonian cannot be regarded as a small perturbation. The resolution of this apparent paradox lies in the fact that $R=2.0$ makes any small disorder a large perturbation. This can be understood as follows. On average, the effect of small disorder is to increase the Euclidean distance between any two sites. For example, if the distance between two sites was initially $a=1$, then it can be easily shown that after distortion, the average distance between all such sites would be $a'\approx 1 + \sigma^2/2$, where $\sigma$ is the standard deviation. As a result, many hoppings that were initially non-zero (that is, within the cut-off radius $R$) are now excluded from the Hamiltonian, leading to a sudden change in the system's connectivity. Therefore, in a square lattice, at any integer value of $R$, even a small disorder can push pairs of sites beyond the cut-off and hence truncating hopping-paths.  On the other hand, when we chose $R=1.5$, all hoppings were between sites that were separated by $1$ unit or $\sqrt{2}$ units in the crystalline system. Consequently, a small distortion only modulated the non-zero matrix elements of the initial Hamiltonian without affecting the entries which were zero. \\

\textbf{Note} :  \cite{alvarez_2022} also discusses the effects of the choice of the hard cut-off in the supplemental material. We thank the referee for bringing our attention to this reference.

\begin{thebibliography}{57}%
\makeatletter
\providecommand \@ifxundefined [1]{%
 \@ifx{#1\undefined}
}%
\providecommand \@ifnum [1]{%
 \ifnum #1\expandafter \@firstoftwo
 \else \expandafter \@secondoftwo
 \fi
}%
\providecommand \@ifx [1]{%
 \ifx #1\expandafter \@firstoftwo
 \else \expandafter \@secondoftwo
 \fi
}%
\providecommand \natexlab [1]{#1}%
\providecommand \enquote  [1]{``#1''}%
\providecommand \bibnamefont  [1]{#1}%
\providecommand \bibfnamefont [1]{#1}%
\providecommand \citenamefont [1]{#1}%
\providecommand \href@noop [0]{\@secondoftwo}%
\providecommand \href [0]{\begingroup \@sanitize@url \@href}%
\providecommand \@href[1]{\@@startlink{#1}\@@href}%
\providecommand \@@href[1]{\endgroup#1\@@endlink}%
\providecommand \@sanitize@url [0]{\catcode `\\12\catcode `\$12\catcode `\&12\catcode `\#12\catcode `\^12\catcode `\_12\catcode `\%12\relax}%
\providecommand \@@startlink[1]{}%
\providecommand \@@endlink[0]{}%
\providecommand \url  [0]{\begingroup\@sanitize@url \@url }%
\providecommand \@url [1]{\endgroup\@href {#1}{\urlprefix }}%
\providecommand \urlprefix  [0]{URL }%
\providecommand \Eprint [0]{\href }%
\providecommand \doibase [0]{https://doi.org/}%
\providecommand \selectlanguage [0]{\@gobble}%
\providecommand \bibinfo  [0]{\@secondoftwo}%
\providecommand \bibfield  [0]{\@secondoftwo}%
\providecommand \translation [1]{[#1]}%
\providecommand \BibitemOpen [0]{}%
\providecommand \bibitemStop [0]{}%
\providecommand \bibitemNoStop [0]{.\EOS\space}%
\providecommand \EOS [0]{\spacefactor3000\relax}%
\providecommand \BibitemShut  [1]{\csname bibitem#1\endcsname}%
\let\auto@bib@innerbib\@empty
\bibitem [{\citenamefont {Hasan}\ and\ \citenamefont {Kane}(2010)}]{hasan}%
  \BibitemOpen
  \bibfield  {author} {\bibinfo {author} {\bibfnamefont {M.~Z.}\ \bibnamefont {Hasan}}\ and\ \bibinfo {author} {\bibfnamefont {C.~L.}\ \bibnamefont {Kane}},\ }\bibfield  {title} {\bibinfo {title} {Colloquium: Topological insulators},\ }\href {https://doi.org/10.1103/RevModPhys.82.3045} {\bibfield  {journal} {\bibinfo  {journal} {Rev. Mod. Phys.}\ }\textbf {\bibinfo {volume} {82}},\ \bibinfo {pages} {3045} (\bibinfo {year} {2010})}\BibitemShut {NoStop}%
\bibitem [{\citenamefont {Qi}\ and\ \citenamefont {Zhang}(2011)}]{qi}%
  \BibitemOpen
  \bibfield  {author} {\bibinfo {author} {\bibfnamefont {X.-L.}\ \bibnamefont {Qi}}\ and\ \bibinfo {author} {\bibfnamefont {S.-C.}\ \bibnamefont {Zhang}},\ }\bibfield  {title} {\bibinfo {title} {Topological insulators and superconductors},\ }\href {https://doi.org/10.1103/RevModPhys.83.1057} {\bibfield  {journal} {\bibinfo  {journal} {Rev. Mod. Phys.}\ }\textbf {\bibinfo {volume} {83}},\ \bibinfo {pages} {1057} (\bibinfo {year} {2011})}\BibitemShut {NoStop}%
\bibitem [{\citenamefont {Schnyder}\ \emph {et~al.}(2008)\citenamefont {Schnyder}, \citenamefont {Ryu}, \citenamefont {Furusaki},\ and\ \citenamefont {Ludwig}}]{schnyder}%
  \BibitemOpen
  \bibfield  {author} {\bibinfo {author} {\bibfnamefont {A.~P.}\ \bibnamefont {Schnyder}}, \bibinfo {author} {\bibfnamefont {S.}~\bibnamefont {Ryu}}, \bibinfo {author} {\bibfnamefont {A.}~\bibnamefont {Furusaki}},\ and\ \bibinfo {author} {\bibfnamefont {A.~W.~W.}\ \bibnamefont {Ludwig}},\ }\bibfield  {title} {\bibinfo {title} {Classification of topological insulators and superconductors in three spatial dimensions},\ }\href {https://doi.org/10.1103/PhysRevB.78.195125} {\bibfield  {journal} {\bibinfo  {journal} {Phys. Rev. B}\ }\textbf {\bibinfo {volume} {78}},\ \bibinfo {pages} {195125} (\bibinfo {year} {2008})}\BibitemShut {NoStop}%
\bibitem [{\citenamefont {Chiu}\ \emph {et~al.}(2016)\citenamefont {Chiu}, \citenamefont {Teo}, \citenamefont {Schnyder},\ and\ \citenamefont {Ryu}}]{Chiu2016}%
  \BibitemOpen
  \bibfield  {author} {\bibinfo {author} {\bibfnamefont {C.-K.}\ \bibnamefont {Chiu}}, \bibinfo {author} {\bibfnamefont {J.~C.~Y.}\ \bibnamefont {Teo}}, \bibinfo {author} {\bibfnamefont {A.~P.}\ \bibnamefont {Schnyder}},\ and\ \bibinfo {author} {\bibfnamefont {S.}~\bibnamefont {Ryu}},\ }\bibfield  {title} {\bibinfo {title} {Classification of topological quantum matter with symmetries},\ }\href {https://doi.org/10.1103/RevModPhys.88.035005} {\bibfield  {journal} {\bibinfo  {journal} {Rev. Mod. Phys.}\ }\textbf {\bibinfo {volume} {88}},\ \bibinfo {pages} {035005} (\bibinfo {year} {2016})}\BibitemShut {NoStop}%
\bibitem [{\citenamefont {Li}\ \emph {et~al.}(2009)\citenamefont {Li}, \citenamefont {Chu}, \citenamefont {Jain},\ and\ \citenamefont {Shen}}]{Li1}%
  \BibitemOpen
  \bibfield  {author} {\bibinfo {author} {\bibfnamefont {J.}~\bibnamefont {Li}}, \bibinfo {author} {\bibfnamefont {R.-L.}\ \bibnamefont {Chu}}, \bibinfo {author} {\bibfnamefont {J.~K.}\ \bibnamefont {Jain}},\ and\ \bibinfo {author} {\bibfnamefont {S.-Q.}\ \bibnamefont {Shen}},\ }\bibfield  {title} {\bibinfo {title} {Topological anderson insulator},\ }\href {https://doi.org/10.1103/PhysRevLett.102.136806} {\bibfield  {journal} {\bibinfo  {journal} {Phys. Rev. Lett.}\ }\textbf {\bibinfo {volume} {102}},\ \bibinfo {pages} {136806} (\bibinfo {year} {2009})}\BibitemShut {NoStop}%
\bibitem [{\citenamefont {N.P.Mithcell}\ \emph {et~al.}(2018)\citenamefont {N.P.Mithcell}, \citenamefont {L.M.Nash}, \citenamefont {D.Hexner}, \citenamefont {A.M.Turner},\ and\ \citenamefont {W.T.M.Irvine}}]{mitchell}%
  \BibitemOpen
  \bibfield  {author} {\bibinfo {author} {\bibnamefont {N.P.Mithcell}}, \bibinfo {author} {\bibnamefont {L.M.Nash}}, \bibinfo {author} {\bibnamefont {D.Hexner}}, \bibinfo {author} {\bibnamefont {A.M.Turner}},\ and\ \bibinfo {author} {\bibnamefont {W.T.M.Irvine}},\ }\bibfield  {title} {\bibinfo {title} {Amorphous topological insulators constructed from random point sets},\ }\href@noop {} {\bibfield  {journal} {\bibinfo  {journal} {Nature Physics}\ }\textbf {\bibinfo {volume} {14}},\ \bibinfo {pages} {380} (\bibinfo {year} {2018})}\BibitemShut {NoStop}%
\bibitem [{\citenamefont {Agarwala}\ and\ \citenamefont {Shenoy}(2017)}]{agarwala1}%
  \BibitemOpen
  \bibfield  {author} {\bibinfo {author} {\bibfnamefont {A.}~\bibnamefont {Agarwala}}\ and\ \bibinfo {author} {\bibfnamefont {V.~B.}\ \bibnamefont {Shenoy}},\ }\bibfield  {title} {\bibinfo {title} {Topological insulators in amorphous systems},\ }\href {https://doi.org/10.1103/PhysRevLett.118.236402} {\bibfield  {journal} {\bibinfo  {journal} {Phys. Rev. Lett.}\ }\textbf {\bibinfo {volume} {118}},\ \bibinfo {pages} {236402} (\bibinfo {year} {2017})}\BibitemShut {NoStop}%
\bibitem [{\citenamefont {Ur\'{\i}a-\'Alvarez}\ and\ \citenamefont {Palacios}(2025)}]{alvarez}%
  \BibitemOpen
  \bibfield  {author} {\bibinfo {author} {\bibfnamefont {A.~J.}\ \bibnamefont {Ur\'{\i}a-\'Alvarez}}\ and\ \bibinfo {author} {\bibfnamefont {J.~J.}\ \bibnamefont {Palacios}},\ }\bibfield  {title} {\bibinfo {title} {Amorphization-induced topological and insulator-metal transitions in bidimensional ${\mathrm{bi}}_{x}{\mathrm{sb}}_{1\ensuremath{-}x}$ alloys},\ }\href {https://doi.org/10.1103/z5tm-cvgn} {\bibfield  {journal} {\bibinfo  {journal} {Phys. Rev. Res.}\ }\textbf {\bibinfo {volume} {7}},\ \bibinfo {pages} {043263} (\bibinfo {year} {2025})}\BibitemShut {NoStop}%
\bibitem [{\citenamefont {Regis}\ \emph {et~al.}(2024)\citenamefont {Regis}, \citenamefont {Velasco}, \citenamefont {Neto},\ and\ \citenamefont {Lewenkopf}}]{regis}%
  \BibitemOpen
  \bibfield  {author} {\bibinfo {author} {\bibfnamefont {V.}~\bibnamefont {Regis}}, \bibinfo {author} {\bibfnamefont {V.}~\bibnamefont {Velasco}}, \bibinfo {author} {\bibfnamefont {M.~B.~S.}\ \bibnamefont {Neto}},\ and\ \bibinfo {author} {\bibfnamefont {C.}~\bibnamefont {Lewenkopf}},\ }\bibfield  {title} {\bibinfo {title} {Structure-driven phase transitions in paracrystalline topological insulators},\ }\href {https://doi.org/10.1103/PhysRevB.110.L161105} {\bibfield  {journal} {\bibinfo  {journal} {Phys. Rev. B}\ }\textbf {\bibinfo {volume} {110}},\ \bibinfo {pages} {L161105} (\bibinfo {year} {2024})}\BibitemShut {NoStop}%
\bibitem [{\citenamefont {Cheng}\ \emph {et~al.}(2023)\citenamefont {Cheng}, \citenamefont {Qu}, \citenamefont {Xiao}, \citenamefont {Jia}, \citenamefont {Chen},\ and\ \citenamefont {Zhang}}]{cheng}%
  \BibitemOpen
  \bibfield  {author} {\bibinfo {author} {\bibfnamefont {X.}~\bibnamefont {Cheng}}, \bibinfo {author} {\bibfnamefont {T.}~\bibnamefont {Qu}}, \bibinfo {author} {\bibfnamefont {L.}~\bibnamefont {Xiao}}, \bibinfo {author} {\bibfnamefont {S.}~\bibnamefont {Jia}}, \bibinfo {author} {\bibfnamefont {J.}~\bibnamefont {Chen}},\ and\ \bibinfo {author} {\bibfnamefont {L.}~\bibnamefont {Zhang}},\ }\bibfield  {title} {\bibinfo {title} {Topological anderson amorphous insulator},\ }\href {https://doi.org/10.1103/PhysRevB.108.L081110} {\bibfield  {journal} {\bibinfo  {journal} {Phys. Rev. B}\ }\textbf {\bibinfo {volume} {108}},\ \bibinfo {pages} {L081110} (\bibinfo {year} {2023})}\BibitemShut {NoStop}%
\bibitem [{\citenamefont {Zhang}\ \emph {et~al.}(2023)\citenamefont {Zhang}, \citenamefont {Delplace},\ and\ \citenamefont {Fleury}}]{fleury1}%
  \BibitemOpen
  \bibfield  {author} {\bibinfo {author} {\bibfnamefont {Z.}~\bibnamefont {Zhang}}, \bibinfo {author} {\bibfnamefont {P.}~\bibnamefont {Delplace}},\ and\ \bibinfo {author} {\bibfnamefont {R.}~\bibnamefont {Fleury}},\ }\bibfield  {title} {\bibinfo {title} {Anomalous topological waves in strongly amorphous scattering networks},\ }\bibfield  {journal} {\bibinfo  {journal} {Science Advances}\ }\textbf {\bibinfo {volume} {9}},\ \href {https://doi.org/10.1126/sciadv.adg3186} {10.1126/sciadv.adg3186} (\bibinfo {year} {2023}),\ \Eprint {https://arxiv.org/abs/https://www.science.org/doi/pdf/10.1126/sciadv.adg3186} {https://www.science.org/doi/pdf/10.1126/sciadv.adg3186} \BibitemShut {NoStop}%
\bibitem [{\citenamefont {Agarwala}\ \emph {et~al.}(2020)\citenamefont {Agarwala}, \citenamefont {Juri\ifmmode \check{c}\else \v{c}\fi{}i\ifmmode~\acute{c}\else \'{c}\fi{}},\ and\ \citenamefont {Roy}}]{agarwala2}%
  \BibitemOpen
  \bibfield  {author} {\bibinfo {author} {\bibfnamefont {A.}~\bibnamefont {Agarwala}}, \bibinfo {author} {\bibfnamefont {V.}~\bibnamefont {Juri\ifmmode \check{c}\else \v{c}\fi{}i\ifmmode~\acute{c}\else \'{c}\fi{}}},\ and\ \bibinfo {author} {\bibfnamefont {B.}~\bibnamefont {Roy}},\ }\bibfield  {title} {\bibinfo {title} {Higher-order topological insulators in amorphous solids},\ }\href {https://doi.org/10.1103/PhysRevResearch.2.012067} {\bibfield  {journal} {\bibinfo  {journal} {Phys. Rev. Res.}\ }\textbf {\bibinfo {volume} {2}},\ \bibinfo {pages} {012067} (\bibinfo {year} {2020})}\BibitemShut {NoStop}%
\bibitem [{\citenamefont {Varjas}\ \emph {et~al.}(2019)\citenamefont {Varjas}, \citenamefont {Lau}, \citenamefont {P\"oyh\"onen}, \citenamefont {Akhmerov}, \citenamefont {Pikulin},\ and\ \citenamefont {Fulga}}]{varjas}%
  \BibitemOpen
  \bibfield  {author} {\bibinfo {author} {\bibfnamefont {D.}~\bibnamefont {Varjas}}, \bibinfo {author} {\bibfnamefont {A.}~\bibnamefont {Lau}}, \bibinfo {author} {\bibfnamefont {K.}~\bibnamefont {P\"oyh\"onen}}, \bibinfo {author} {\bibfnamefont {A.~R.}\ \bibnamefont {Akhmerov}}, \bibinfo {author} {\bibfnamefont {D.~I.}\ \bibnamefont {Pikulin}},\ and\ \bibinfo {author} {\bibfnamefont {I.~C.}\ \bibnamefont {Fulga}},\ }\bibfield  {title} {\bibinfo {title} {Topological phases without crystalline counterparts},\ }\href {https://doi.org/10.1103/PhysRevLett.123.196401} {\bibfield  {journal} {\bibinfo  {journal} {Phys. Rev. Lett.}\ }\textbf {\bibinfo {volume} {123}},\ \bibinfo {pages} {196401} (\bibinfo {year} {2019})}\BibitemShut {NoStop}%
\bibitem [{\citenamefont {Grushin}\ and\ \citenamefont {Repellin}(2023)}]{grushin}%
  \BibitemOpen
  \bibfield  {author} {\bibinfo {author} {\bibfnamefont {A.~G.}\ \bibnamefont {Grushin}}\ and\ \bibinfo {author} {\bibfnamefont {C.}~\bibnamefont {Repellin}},\ }\bibfield  {title} {\bibinfo {title} {Amorphous and polycrystalline routes toward a chiral spin liquid},\ }\href {https://doi.org/10.1103/PhysRevLett.130.186702} {\bibfield  {journal} {\bibinfo  {journal} {Phys. Rev. Lett.}\ }\textbf {\bibinfo {volume} {130}},\ \bibinfo {pages} {186702} (\bibinfo {year} {2023})}\BibitemShut {NoStop}%
\bibitem [{\citenamefont {P.Corbae}\ \emph {et~al.}(2023)\citenamefont {P.Corbae}, \citenamefont {J.D.Hannukainen}, \citenamefont {Q.Marsal}, \citenamefont {D.M.-Segovia},\ and\ \citenamefont {A.G.Grushin}}]{corbae-review}%
  \BibitemOpen
  \bibfield  {author} {\bibinfo {author} {\bibnamefont {P.Corbae}}, \bibinfo {author} {\bibnamefont {J.D.Hannukainen}}, \bibinfo {author} {\bibnamefont {Q.Marsal}}, \bibinfo {author} {\bibnamefont {D.M.-Segovia}},\ and\ \bibinfo {author} {\bibnamefont {A.G.Grushin}},\ }\bibfield  {title} {\bibinfo {title} {Amorphous topological matter: theory and experiment},\ }\href@noop {} {\bibfield  {journal} {\bibinfo  {journal} {EPL}\ }\textbf {\bibinfo {volume} {142}},\ \bibinfo {pages} {16001} (\bibinfo {year} {2023})}\BibitemShut {NoStop}%
\bibitem [{\citenamefont {Marsal}\ \emph {et~al.}(2020)\citenamefont {Marsal}, \citenamefont {Varjas},\ and\ \citenamefont {Grushin}}]{marsal}%
  \BibitemOpen
  \bibfield  {author} {\bibinfo {author} {\bibfnamefont {Q.}~\bibnamefont {Marsal}}, \bibinfo {author} {\bibfnamefont {D.}~\bibnamefont {Varjas}},\ and\ \bibinfo {author} {\bibfnamefont {A.~G.}\ \bibnamefont {Grushin}},\ }\bibfield  {title} {\bibinfo {title} {Topological weaire–thorpe models of amorphous matter},\ }\href {https://doi.org/10.1073/pnas.2007384117} {\bibfield  {journal} {\bibinfo  {journal} {Proceedings of the National Academy of Sciences}\ }\textbf {\bibinfo {volume} {117}},\ \bibinfo {pages} {30260} (\bibinfo {year} {2020})},\ \Eprint {https://arxiv.org/abs/https://www.pnas.org/doi/pdf/10.1073/pnas.2007384117} {https://www.pnas.org/doi/pdf/10.1073/pnas.2007384117} \BibitemShut {NoStop}%
\bibitem [{\citenamefont {Cassella}\ \emph {et~al.}(2023)\citenamefont {Cassella}, \citenamefont {d'Ornellas}, \citenamefont {Hodson}, \citenamefont {Natori},\ and\ \citenamefont {Knolle}}]{Cassella2023-nq}%
  \BibitemOpen
  \bibfield  {author} {\bibinfo {author} {\bibfnamefont {G.}~\bibnamefont {Cassella}}, \bibinfo {author} {\bibfnamefont {P.}~\bibnamefont {d'Ornellas}}, \bibinfo {author} {\bibfnamefont {T.}~\bibnamefont {Hodson}}, \bibinfo {author} {\bibfnamefont {W.~M.~H.}\ \bibnamefont {Natori}},\ and\ \bibinfo {author} {\bibfnamefont {J.}~\bibnamefont {Knolle}},\ }\bibfield  {title} {\bibinfo {title} {An exact chiral amorphous spin liquid},\ }\href@noop {} {\bibfield  {journal} {\bibinfo  {journal} {Nature Communications}\ }\textbf {\bibinfo {volume} {14}},\ \bibinfo {pages} {6663} (\bibinfo {year} {2023})}\BibitemShut {NoStop}%
\bibitem [{\citenamefont {Yang}\ \emph {et~al.}(2019)\citenamefont {Yang}, \citenamefont {Qin}, \citenamefont {Deng}, \citenamefont {Duan},\ and\ \citenamefont {Xu}}]{yang}%
  \BibitemOpen
  \bibfield  {author} {\bibinfo {author} {\bibfnamefont {Y.-B.}\ \bibnamefont {Yang}}, \bibinfo {author} {\bibfnamefont {T.}~\bibnamefont {Qin}}, \bibinfo {author} {\bibfnamefont {D.-L.}\ \bibnamefont {Deng}}, \bibinfo {author} {\bibfnamefont {L.-M.}\ \bibnamefont {Duan}},\ and\ \bibinfo {author} {\bibfnamefont {Y.}~\bibnamefont {Xu}},\ }\bibfield  {title} {\bibinfo {title} {Topological amorphous metals},\ }\href {https://doi.org/10.1103/PhysRevLett.123.076401} {\bibfield  {journal} {\bibinfo  {journal} {Phys. Rev. Lett.}\ }\textbf {\bibinfo {volume} {123}},\ \bibinfo {pages} {076401} (\bibinfo {year} {2019})}\BibitemShut {NoStop}%
\bibitem [{\citenamefont {Mansha}\ and\ \citenamefont {Chong}(2017)}]{mansha}%
  \BibitemOpen
  \bibfield  {author} {\bibinfo {author} {\bibfnamefont {S.}~\bibnamefont {Mansha}}\ and\ \bibinfo {author} {\bibfnamefont {Y.~D.}\ \bibnamefont {Chong}},\ }\bibfield  {title} {\bibinfo {title} {Robust edge states in amorphous gyromagnetic photonic lattices},\ }\href {https://doi.org/10.1103/PhysRevB.96.121405} {\bibfield  {journal} {\bibinfo  {journal} {Phys. Rev. B}\ }\textbf {\bibinfo {volume} {96}},\ \bibinfo {pages} {121405} (\bibinfo {year} {2017})}\BibitemShut {NoStop}%
\bibitem [{\citenamefont {Costa}\ \emph {et~al.}(2019)\citenamefont {Costa}, \citenamefont {Schleder}, \citenamefont {Buongiorno~Nardelli}, \citenamefont {Lewenkopf},\ and\ \citenamefont {Fazzio}}]{costa}%
  \BibitemOpen
  \bibfield  {author} {\bibinfo {author} {\bibfnamefont {M.}~\bibnamefont {Costa}}, \bibinfo {author} {\bibfnamefont {G.~R.}\ \bibnamefont {Schleder}}, \bibinfo {author} {\bibfnamefont {M.}~\bibnamefont {Buongiorno~Nardelli}}, \bibinfo {author} {\bibfnamefont {C.}~\bibnamefont {Lewenkopf}},\ and\ \bibinfo {author} {\bibfnamefont {A.}~\bibnamefont {Fazzio}},\ }\bibfield  {title} {\bibinfo {title} {Toward realistic amorphous topological insulators},\ }\href {https://doi.org/10.1021/acs.nanolett.9b03881} {\bibfield  {journal} {\bibinfo  {journal} {Nano Letters}\ }\textbf {\bibinfo {volume} {19}},\ \bibinfo {pages} {8941} (\bibinfo {year} {2019})},\ \bibinfo {note} {pMID: 31679336},\ \Eprint {https://arxiv.org/abs/https://doi.org/10.1021/acs.nanolett.9b03881} {https://doi.org/10.1021/acs.nanolett.9b03881} \BibitemShut {NoStop}%
\bibitem [{\citenamefont {Spring}\ \emph {et~al.}(2021)\citenamefont {Spring}, \citenamefont {Akhmerov},\ and\ \citenamefont {Varjas}}]{spring1}%
  \BibitemOpen
  \bibfield  {author} {\bibinfo {author} {\bibfnamefont {H.}~\bibnamefont {Spring}}, \bibinfo {author} {\bibfnamefont {A.~R.}\ \bibnamefont {Akhmerov}},\ and\ \bibinfo {author} {\bibfnamefont {D.}~\bibnamefont {Varjas}},\ }\bibfield  {title} {\bibinfo {title} {{Amorphous topological phases protected by continuous rotation symmetry}},\ }\href {https://doi.org/10.21468/SciPostPhys.11.2.022} {\bibfield  {journal} {\bibinfo  {journal} {SciPost Phys.}\ }\textbf {\bibinfo {volume} {11}},\ \bibinfo {pages} {022} (\bibinfo {year} {2021})}\BibitemShut {NoStop}%
\bibitem [{\citenamefont {Kim}\ \emph {et~al.}(2023)\citenamefont {Kim}, \citenamefont {Agarwala},\ and\ \citenamefont {Chowdhury}}]{kim}%
  \BibitemOpen
  \bibfield  {author} {\bibinfo {author} {\bibfnamefont {S.}~\bibnamefont {Kim}}, \bibinfo {author} {\bibfnamefont {A.}~\bibnamefont {Agarwala}},\ and\ \bibinfo {author} {\bibfnamefont {D.}~\bibnamefont {Chowdhury}},\ }\bibfield  {title} {\bibinfo {title} {Fractionalization and topology in amorphous electronic solids},\ }\href {https://doi.org/10.1103/PhysRevLett.130.026202} {\bibfield  {journal} {\bibinfo  {journal} {Phys. Rev. Lett.}\ }\textbf {\bibinfo {volume} {130}},\ \bibinfo {pages} {026202} (\bibinfo {year} {2023})}\BibitemShut {NoStop}%
\bibitem [{\citenamefont {Manna}\ \emph {et~al.}(2024)\citenamefont {Manna}, \citenamefont {Das},\ and\ \citenamefont {Roy}}]{manna}%
  \BibitemOpen
  \bibfield  {author} {\bibinfo {author} {\bibfnamefont {S.}~\bibnamefont {Manna}}, \bibinfo {author} {\bibfnamefont {S.~K.}\ \bibnamefont {Das}},\ and\ \bibinfo {author} {\bibfnamefont {B.}~\bibnamefont {Roy}},\ }\bibfield  {title} {\bibinfo {title} {Noncrystalline topological superconductors},\ }\href {https://doi.org/10.1103/PhysRevB.109.174512} {\bibfield  {journal} {\bibinfo  {journal} {Phys. Rev. B}\ }\textbf {\bibinfo {volume} {109}},\ \bibinfo {pages} {174512} (\bibinfo {year} {2024})}\BibitemShut {NoStop}%
\bibitem [{\citenamefont {Martínez}\ \emph {et~al.}(2026)\citenamefont {Martínez}, \citenamefont {Jezequel}, \citenamefont {Bardarson}, \citenamefont {Kvorning},\ and\ \citenamefont {Hannukainen}}]{martinez2025}%
  \BibitemOpen
  \bibfield  {author} {\bibinfo {author} {\bibfnamefont {M.~F.}\ \bibnamefont {Martínez}}, \bibinfo {author} {\bibfnamefont {L.}~\bibnamefont {Jezequel}}, \bibinfo {author} {\bibfnamefont {J.~H.}\ \bibnamefont {Bardarson}}, \bibinfo {author} {\bibfnamefont {T.~K.}\ \bibnamefont {Kvorning}},\ and\ \bibinfo {author} {\bibfnamefont {J.~D.}\ \bibnamefont {Hannukainen}},\ }\bibfield  {title} {\bibinfo {title} {One-particle density matrix framework for mode-shell correspondence: Characterizing topology in amorphous higher-order topological insulators},\ }\href {https://doi.org/10.1103/rt48-gpfm} {\bibfield  {journal} {\bibinfo  {journal} {Phys. Rev. Res.}\ } (\bibinfo {year} {2026})}\BibitemShut {NoStop}%
\bibitem [{\citenamefont {Ghosh}\ \emph {et~al.}(2026)\citenamefont {Ghosh}, \citenamefont {Marsal},\ and\ \citenamefont {Black-Schaffer}}]{ghosh2025}%
  \BibitemOpen
  \bibfield  {author} {\bibinfo {author} {\bibfnamefont {A.~K.}\ \bibnamefont {Ghosh}}, \bibinfo {author} {\bibfnamefont {Q.}~\bibnamefont {Marsal}},\ and\ \bibinfo {author} {\bibfnamefont {A.~M.}\ \bibnamefont {Black-Schaffer}},\ }\bibfield  {title} {\bibinfo {title} {Laser-induced topological phases in monolayer amorphous carbon},\ }\bibfield  {journal} {\bibinfo  {journal} {Physical Review B}\ }\textbf {\bibinfo {volume} {113}},\ \href {https://doi.org/10.1103/tp7r-31dj} {10.1103/tp7r-31dj} (\bibinfo {year} {2026})\BibitemShut {NoStop}%
\bibitem [{\citenamefont {Jezequel}\ \emph {et~al.}(2026)\citenamefont {Jezequel}, \citenamefont {Bardarson},\ and\ \citenamefont {Grushin}}]{jezequel2025}%
  \BibitemOpen
  \bibfield  {author} {\bibinfo {author} {\bibfnamefont {L.}~\bibnamefont {Jezequel}}, \bibinfo {author} {\bibfnamefont {J.~H.}\ \bibnamefont {Bardarson}},\ and\ \bibinfo {author} {\bibfnamefont {A.~G.}\ \bibnamefont {Grushin}},\ }\bibfield  {title} {\bibinfo {title} {Explicit equivalence between the spectral localizer and local chern and winding markers},\ }\bibfield  {journal} {\bibinfo  {journal} {SciPost Physics}\ }\textbf {\bibinfo {volume} {20}},\ \href {https://doi.org/10.21468/scipostphys.20.4.118} {10.21468/scipostphys.20.4.118} (\bibinfo {year} {2026})\BibitemShut {NoStop}%
\bibitem [{\citenamefont {Corbae}\ \emph {et~al.}(2023)\citenamefont {Corbae}, \citenamefont {Ciocys}, \citenamefont {Varjas}, \citenamefont {Kennedy}, \citenamefont {Zeltmann}, \citenamefont {Molina-Ruiz}, \citenamefont {Griffin}, \citenamefont {Jozwiak}, \citenamefont {Chen}, \citenamefont {Wang}, \citenamefont {Minor}, \citenamefont {Scott}, \citenamefont {Grushin}, \citenamefont {Lanzara},\ and\ \citenamefont {Hellman}}]{Corbae2023-kq}%
  \BibitemOpen
  \bibfield  {author} {\bibinfo {author} {\bibfnamefont {P.}~\bibnamefont {Corbae}}, \bibinfo {author} {\bibfnamefont {S.}~\bibnamefont {Ciocys}}, \bibinfo {author} {\bibfnamefont {D.}~\bibnamefont {Varjas}}, \bibinfo {author} {\bibfnamefont {E.}~\bibnamefont {Kennedy}}, \bibinfo {author} {\bibfnamefont {S.}~\bibnamefont {Zeltmann}}, \bibinfo {author} {\bibfnamefont {M.}~\bibnamefont {Molina-Ruiz}}, \bibinfo {author} {\bibfnamefont {S.~M.}\ \bibnamefont {Griffin}}, \bibinfo {author} {\bibfnamefont {C.}~\bibnamefont {Jozwiak}}, \bibinfo {author} {\bibfnamefont {Z.}~\bibnamefont {Chen}}, \bibinfo {author} {\bibfnamefont {L.-W.}\ \bibnamefont {Wang}}, \bibinfo {author} {\bibfnamefont {A.~M.}\ \bibnamefont {Minor}}, \bibinfo {author} {\bibfnamefont {M.}~\bibnamefont {Scott}}, \bibinfo {author} {\bibfnamefont {A.~G.}\ \bibnamefont {Grushin}}, \bibinfo {author} {\bibfnamefont {A.}~\bibnamefont {Lanzara}},\ and\ \bibinfo {author} {\bibfnamefont {F.}~\bibnamefont {Hellman}},\ }\bibfield  {title} {\bibinfo {title}
  {Observation of spin-momentum locked surface states in amorphous {Bi2Se3}},\ }\href@noop {} {\bibfield  {journal} {\bibinfo  {journal} {Nature Materials}\ }\textbf {\bibinfo {volume} {22}},\ \bibinfo {pages} {200} (\bibinfo {year} {2023})}\BibitemShut {NoStop}%
\bibitem [{\citenamefont {Ciocys}\ \emph {et~al.}(2024)\citenamefont {Ciocys}, \citenamefont {Marsal}, \citenamefont {Corbae}, \citenamefont {Varjas}, \citenamefont {Kennedy}, \citenamefont {Scott}, \citenamefont {Hellman}, \citenamefont {Grushin},\ and\ \citenamefont {Lanzara}}]{Ciocys2024}%
  \BibitemOpen
  \bibfield  {author} {\bibinfo {author} {\bibfnamefont {S.~T.}\ \bibnamefont {Ciocys}}, \bibinfo {author} {\bibfnamefont {Q.}~\bibnamefont {Marsal}}, \bibinfo {author} {\bibfnamefont {P.}~\bibnamefont {Corbae}}, \bibinfo {author} {\bibfnamefont {D.}~\bibnamefont {Varjas}}, \bibinfo {author} {\bibfnamefont {E.}~\bibnamefont {Kennedy}}, \bibinfo {author} {\bibfnamefont {M.}~\bibnamefont {Scott}}, \bibinfo {author} {\bibfnamefont {F.}~\bibnamefont {Hellman}}, \bibinfo {author} {\bibfnamefont {A.~G.}\ \bibnamefont {Grushin}},\ and\ \bibinfo {author} {\bibfnamefont {A.}~\bibnamefont {Lanzara}},\ }\bibfield  {title} {\bibinfo {title} {Establishing coherent momentum-space electronic states in locally ordered materials},\ }\href@noop {} {\bibfield  {journal} {\bibinfo  {journal} {Nature Communications}\ }\textbf {\bibinfo {volume} {15}},\ \bibinfo {pages} {8141} (\bibinfo {year} {2024})}\BibitemShut {NoStop}%
\bibitem [{\citenamefont {Zhou}\ \emph {et~al.}(2020)\citenamefont {Zhou}, \citenamefont {Liu}, \citenamefont {Ren}, \citenamefont {Yang}, \citenamefont {Xue}, \citenamefont {Bi}, \citenamefont {Deng}, \citenamefont {Chong},\ and\ \citenamefont {Zhang}}]{zhou2020}%
  \BibitemOpen
  \bibfield  {author} {\bibinfo {author} {\bibfnamefont {P.}~\bibnamefont {Zhou}}, \bibinfo {author} {\bibfnamefont {G.-G.}\ \bibnamefont {Liu}}, \bibinfo {author} {\bibfnamefont {X.}~\bibnamefont {Ren}}, \bibinfo {author} {\bibfnamefont {Y.}~\bibnamefont {Yang}}, \bibinfo {author} {\bibfnamefont {H.}~\bibnamefont {Xue}}, \bibinfo {author} {\bibfnamefont {L.}~\bibnamefont {Bi}}, \bibinfo {author} {\bibfnamefont {L.}~\bibnamefont {Deng}}, \bibinfo {author} {\bibfnamefont {Y.}~\bibnamefont {Chong}},\ and\ \bibinfo {author} {\bibfnamefont {B.}~\bibnamefont {Zhang}},\ }\bibfield  {title} {\bibinfo {title} {Photonic amorphous topological insulator},\ }\href@noop {} {\bibfield  {journal} {\bibinfo  {journal} {Light: Science \& Applications}\ }\textbf {\bibinfo {volume} {9}},\ \bibinfo {pages} {133} (\bibinfo {year} {2020})}\BibitemShut {NoStop}%
\bibitem [{\citenamefont {Rechtsman}\ \emph {et~al.}(2011)\citenamefont {Rechtsman}, \citenamefont {Szameit}, \citenamefont {Dreisow}, \citenamefont {Heinrich}, \citenamefont {Keil}, \citenamefont {Nolte},\ and\ \citenamefont {Segev}}]{rechtsman}%
  \BibitemOpen
  \bibfield  {author} {\bibinfo {author} {\bibfnamefont {M.}~\bibnamefont {Rechtsman}}, \bibinfo {author} {\bibfnamefont {A.}~\bibnamefont {Szameit}}, \bibinfo {author} {\bibfnamefont {F.}~\bibnamefont {Dreisow}}, \bibinfo {author} {\bibfnamefont {M.}~\bibnamefont {Heinrich}}, \bibinfo {author} {\bibfnamefont {R.}~\bibnamefont {Keil}}, \bibinfo {author} {\bibfnamefont {S.}~\bibnamefont {Nolte}},\ and\ \bibinfo {author} {\bibfnamefont {M.}~\bibnamefont {Segev}},\ }\bibfield  {title} {\bibinfo {title} {Amorphous photonic lattices: Band gaps, effective mass, and suppressed transport},\ }\href {https://doi.org/10.1103/PhysRevLett.106.193904} {\bibfield  {journal} {\bibinfo  {journal} {Phys. Rev. Lett.}\ }\textbf {\bibinfo {volume} {106}},\ \bibinfo {pages} {193904} (\bibinfo {year} {2011})}\BibitemShut {NoStop}%
\bibitem [{\citenamefont {St{\"u}tzer}\ \emph {et~al.}(2018)\citenamefont {St{\"u}tzer}, \citenamefont {Plotnik}, \citenamefont {Lumer}, \citenamefont {Titum}, \citenamefont {Lindner}, \citenamefont {Segev}, \citenamefont {Rechtsman},\ and\ \citenamefont {Szameit}}]{stutzer}%
  \BibitemOpen
  \bibfield  {author} {\bibinfo {author} {\bibfnamefont {S.}~\bibnamefont {St{\"u}tzer}}, \bibinfo {author} {\bibfnamefont {Y.}~\bibnamefont {Plotnik}}, \bibinfo {author} {\bibfnamefont {Y.}~\bibnamefont {Lumer}}, \bibinfo {author} {\bibfnamefont {P.}~\bibnamefont {Titum}}, \bibinfo {author} {\bibfnamefont {N.~H.}\ \bibnamefont {Lindner}}, \bibinfo {author} {\bibfnamefont {M.}~\bibnamefont {Segev}}, \bibinfo {author} {\bibfnamefont {M.~C.}\ \bibnamefont {Rechtsman}},\ and\ \bibinfo {author} {\bibfnamefont {A.}~\bibnamefont {Szameit}},\ }\bibfield  {title} {\bibinfo {title} {Photonic topological anderson insulators},\ }\href@noop {} {\bibfield  {journal} {\bibinfo  {journal} {Nature}\ }\textbf {\bibinfo {volume} {560}},\ \bibinfo {pages} {461} (\bibinfo {year} {2018})}\BibitemShut {NoStop}%
\bibitem [{\citenamefont {Groth}\ \emph {et~al.}(2009)\citenamefont {Groth}, \citenamefont {Wimmer}, \citenamefont {Akhmerov}, \citenamefont {Tworzyd\l{}o},\ and\ \citenamefont {Beenakker}}]{groth2}%
  \BibitemOpen
  \bibfield  {author} {\bibinfo {author} {\bibfnamefont {C.~W.}\ \bibnamefont {Groth}}, \bibinfo {author} {\bibfnamefont {M.}~\bibnamefont {Wimmer}}, \bibinfo {author} {\bibfnamefont {A.~R.}\ \bibnamefont {Akhmerov}}, \bibinfo {author} {\bibfnamefont {J.}~\bibnamefont {Tworzyd\l{}o}},\ and\ \bibinfo {author} {\bibfnamefont {C.~W.~J.}\ \bibnamefont {Beenakker}},\ }\bibfield  {title} {\bibinfo {title} {Theory of the topological anderson insulator},\ }\href {https://doi.org/10.1103/PhysRevLett.103.196805} {\bibfield  {journal} {\bibinfo  {journal} {Phys. Rev. Lett.}\ }\textbf {\bibinfo {volume} {103}},\ \bibinfo {pages} {196805} (\bibinfo {year} {2009})}\BibitemShut {NoStop}%
\bibitem [{\citenamefont {Orth}\ \emph {et~al.}(2016)\citenamefont {Orth}, \citenamefont {Sekera}, \citenamefont {Bruder},\ and\ \citenamefont {Schmidt}}]{orth2016}%
  \BibitemOpen
  \bibfield  {author} {\bibinfo {author} {\bibfnamefont {C.~P.}\ \bibnamefont {Orth}}, \bibinfo {author} {\bibfnamefont {T.}~\bibnamefont {Sekera}}, \bibinfo {author} {\bibfnamefont {C.}~\bibnamefont {Bruder}},\ and\ \bibinfo {author} {\bibfnamefont {T.~L.}\ \bibnamefont {Schmidt}},\ }\bibfield  {title} {\bibinfo {title} {The topological anderson insulator phase in the {Kane-Mele} model},\ }\href@noop {} {\bibfield  {journal} {\bibinfo  {journal} {Scientific Reports}\ }\textbf {\bibinfo {volume} {6}},\ \bibinfo {pages} {24007} (\bibinfo {year} {2016})}\BibitemShut {NoStop}%
\bibitem [{\citenamefont {Song}\ \emph {et~al.}(2012)\citenamefont {Song}, \citenamefont {Liu}, \citenamefont {Jiang}, \citenamefont {Sun},\ and\ \citenamefont {Xie}}]{song1}%
  \BibitemOpen
  \bibfield  {author} {\bibinfo {author} {\bibfnamefont {J.}~\bibnamefont {Song}}, \bibinfo {author} {\bibfnamefont {H.}~\bibnamefont {Liu}}, \bibinfo {author} {\bibfnamefont {H.}~\bibnamefont {Jiang}}, \bibinfo {author} {\bibfnamefont {Q.-f.}\ \bibnamefont {Sun}},\ and\ \bibinfo {author} {\bibfnamefont {X.~C.}\ \bibnamefont {Xie}},\ }\bibfield  {title} {\bibinfo {title} {Dependence of topological anderson insulator on the type of disorder},\ }\href {https://doi.org/10.1103/PhysRevB.85.195125} {\bibfield  {journal} {\bibinfo  {journal} {Phys. Rev. B}\ }\textbf {\bibinfo {volume} {85}},\ \bibinfo {pages} {195125} (\bibinfo {year} {2012})}\BibitemShut {NoStop}%
\bibitem [{\citenamefont {Bernevig}\ \emph {et~al.}(2006)\citenamefont {Bernevig}, \citenamefont {T.Hughes},\ and\ \citenamefont {S.\-Cheng.Zhang}}]{bernevig-science}%
  \BibitemOpen
  \bibfield  {author} {\bibinfo {author} {\bibfnamefont {B.}~\bibnamefont {Bernevig}}, \bibinfo {author} {\bibnamefont {T.Hughes}},\ and\ \bibinfo {author} {\bibnamefont {S.\-Cheng.Zhang}},\ }\bibfield  {title} {\bibinfo {title} {Quantum spin hall effect and topological phase transition in hgte quantum wells},\ }\href@noop {} {\bibfield  {journal} {\bibinfo  {journal} {Nat. Mat.}\ }\textbf {\bibinfo {volume} {314}},\ \bibinfo {pages} {1757} (\bibinfo {year} {2006})}\BibitemShut {NoStop}%
\bibitem [{\citenamefont {Chen}(2023)}]{wchen1}%
  \BibitemOpen
  \bibfield  {author} {\bibinfo {author} {\bibfnamefont {W.}~\bibnamefont {Chen}},\ }\bibfield  {title} {\bibinfo {title} {Universal topological marker},\ }\href {https://doi.org/10.1103/PhysRevB.107.045111} {\bibfield  {journal} {\bibinfo  {journal} {Phys. Rev. B}\ }\textbf {\bibinfo {volume} {107}},\ \bibinfo {pages} {045111} (\bibinfo {year} {2023})}\BibitemShut {NoStop}%
\bibitem [{\citenamefont {Oliveira}\ and\ \citenamefont {Chen}(2024)}]{oliveira}%
  \BibitemOpen
  \bibfield  {author} {\bibinfo {author} {\bibfnamefont {L.~A.}\ \bibnamefont {Oliveira}}\ and\ \bibinfo {author} {\bibfnamefont {W.}~\bibnamefont {Chen}},\ }\bibfield  {title} {\bibinfo {title} {Robustness of topological order against disorder},\ }\href {https://doi.org/10.1103/PhysRevB.109.094202} {\bibfield  {journal} {\bibinfo  {journal} {Phys. Rev. B}\ }\textbf {\bibinfo {volume} {109}},\ \bibinfo {pages} {094202} (\bibinfo {year} {2024})}\BibitemShut {NoStop}%
\bibitem [{\citenamefont {B.A.Bernevig}\ \emph {et~al.}(2006)\citenamefont {B.A.Bernevig}, \citenamefont {T.L.Hughes},\ and\ \citenamefont {S.\-C.Zhang}}]{bernevig1}%
  \BibitemOpen
  \bibfield  {author} {\bibinfo {author} {\bibnamefont {B.A.Bernevig}}, \bibinfo {author} {\bibnamefont {T.L.Hughes}},\ and\ \bibinfo {author} {\bibnamefont {S.\-C.Zhang}},\ }\bibfield  {title} {\bibinfo {title} {Quantum spin hall effect and topological phase transition in hgte quantum wells},\ }\href {https://doi.org/10.1126/science.1133734} {\bibfield  {journal} {\bibinfo  {journal} {Science}\ }\textbf {\bibinfo {volume} {314}},\ \bibinfo {pages} {1757} (\bibinfo {year} {2006})},\ \Eprint {https://arxiv.org/abs/https://www.science.org/doi/pdf/10.1126/science.1133734} {https://www.science.org/doi/pdf/10.1126/science.1133734} \BibitemShut {NoStop}%
\bibitem [{\citenamefont {Konig}\ \emph {et~al.}(2007)\citenamefont {Konig}, \citenamefont {Wiedmann}, \citenamefont {Brüne}, \citenamefont {Roth}, \citenamefont {Buhmann}, \citenamefont {Molenkamp}, \citenamefont {Qi},\ and\ \citenamefont {Zhang}}]{Konig}%
  \BibitemOpen
  \bibfield  {author} {\bibinfo {author} {\bibfnamefont {M.}~\bibnamefont {Konig}}, \bibinfo {author} {\bibfnamefont {S.}~\bibnamefont {Wiedmann}}, \bibinfo {author} {\bibfnamefont {C.}~\bibnamefont {Brüne}}, \bibinfo {author} {\bibfnamefont {A.}~\bibnamefont {Roth}}, \bibinfo {author} {\bibfnamefont {H.}~\bibnamefont {Buhmann}}, \bibinfo {author} {\bibfnamefont {L.~W.}\ \bibnamefont {Molenkamp}}, \bibinfo {author} {\bibfnamefont {X.-L.}\ \bibnamefont {Qi}},\ and\ \bibinfo {author} {\bibfnamefont {S.-C.}\ \bibnamefont {Zhang}},\ }\bibfield  {title} {\bibinfo {title} {Quantum spin hall insulator state in hgte quantum wells},\ }\href {https://doi.org/10.1126/science.1148047} {\bibfield  {journal} {\bibinfo  {journal} {Science}\ }\textbf {\bibinfo {volume} {318}},\ \bibinfo {pages} {766–770} (\bibinfo {year} {2007})}\BibitemShut {NoStop}%
\bibitem [{\citenamefont {Chen}(2020)}]{wchen2}%
  \BibitemOpen
  \bibfield  {author} {\bibinfo {author} {\bibfnamefont {W.}~\bibnamefont {Chen}},\ }\bibfield  {title} {\bibinfo {title} {Absence of equilibrium edge currents in theoretical models of topological insulators},\ }\href {https://doi.org/10.1103/PhysRevB.101.195120} {\bibfield  {journal} {\bibinfo  {journal} {Phys. Rev. B}\ }\textbf {\bibinfo {volume} {101}},\ \bibinfo {pages} {195120} (\bibinfo {year} {2020})}\BibitemShut {NoStop}%
\bibitem [{\citenamefont {Wang}\ \emph {et~al.}(2022)\citenamefont {Wang}, \citenamefont {Cheng}, \citenamefont {Liu}, \citenamefont {Liu},\ and\ \citenamefont {Huang}}]{citian-wang}%
  \BibitemOpen
  \bibfield  {author} {\bibinfo {author} {\bibfnamefont {C.}~\bibnamefont {Wang}}, \bibinfo {author} {\bibfnamefont {T.}~\bibnamefont {Cheng}}, \bibinfo {author} {\bibfnamefont {Z.}~\bibnamefont {Liu}}, \bibinfo {author} {\bibfnamefont {F.}~\bibnamefont {Liu}},\ and\ \bibinfo {author} {\bibfnamefont {H.}~\bibnamefont {Huang}},\ }\bibfield  {title} {\bibinfo {title} {Structural amorphization-induced topological order},\ }\href {https://doi.org/10.1103/PhysRevLett.128.056401} {\bibfield  {journal} {\bibinfo  {journal} {Phys. Rev. Lett.}\ }\textbf {\bibinfo {volume} {128}},\ \bibinfo {pages} {056401} (\bibinfo {year} {2022})}\BibitemShut {NoStop}%
\bibitem [{\citenamefont {Franca}\ and\ \citenamefont {Grushin}(2024)}]{franca1}%
  \BibitemOpen
  \bibfield  {author} {\bibinfo {author} {\bibfnamefont {S.}~\bibnamefont {Franca}}\ and\ \bibinfo {author} {\bibfnamefont {A.~G.}\ \bibnamefont {Grushin}},\ }\bibfield  {title} {\bibinfo {title} {Topological diffusive metal in amorphous transition metal monosilicides},\ }\href {https://doi.org/10.1103/PhysRevMaterials.8.L021201} {\bibfield  {journal} {\bibinfo  {journal} {Phys. Rev. Mater.}\ }\textbf {\bibinfo {volume} {8}},\ \bibinfo {pages} {L021201} (\bibinfo {year} {2024})}\BibitemShut {NoStop}%
\bibitem [{\citenamefont {Bernevig}\ and\ \citenamefont {Hughes}(2013)}]{2963}%
  \BibitemOpen
  \bibfield  {author} {\bibinfo {author} {\bibfnamefont {B.}~\bibnamefont {Bernevig}}\ and\ \bibinfo {author} {\bibfnamefont {T.}~\bibnamefont {Hughes}},\ }\href@noop {} {\emph {\bibinfo {title} {Topological insulators and topological superconductors}}}\ (\bibinfo  {publisher} {Princeton Uni. Press},\ \bibinfo {address} {Princeton},\ \bibinfo {year} {2013})\BibitemShut {NoStop}%
\bibitem [{\citenamefont {Fu}\ and\ \citenamefont {Kane}(2007)}]{fu}%
  \BibitemOpen
  \bibfield  {author} {\bibinfo {author} {\bibfnamefont {L.}~\bibnamefont {Fu}}\ and\ \bibinfo {author} {\bibfnamefont {C.~L.}\ \bibnamefont {Kane}},\ }\bibfield  {title} {\bibinfo {title} {Topological insulators with inversion symmetry},\ }\href {https://doi.org/10.1103/PhysRevB.76.045302} {\bibfield  {journal} {\bibinfo  {journal} {Phys. Rev. B}\ }\textbf {\bibinfo {volume} {76}},\ \bibinfo {pages} {045302} (\bibinfo {year} {2007})}\BibitemShut {NoStop}%
\bibitem [{\citenamefont {Kitaev}(2006)}]{kitaev}%
  \BibitemOpen
  \bibfield  {author} {\bibinfo {author} {\bibfnamefont {A.}~\bibnamefont {Kitaev}},\ }\bibfield  {title} {\bibinfo {title} {Anyons in an exactly solved model and beyond},\ }\href {https://doi.org/https://doi.org/10.1016/j.aop.2005.10.005} {\bibfield  {journal} {\bibinfo  {journal} {Annals of Physics}\ }\textbf {\bibinfo {volume} {321}},\ \bibinfo {pages} {2} (\bibinfo {year} {2006})},\ \bibinfo {note} {january Special Issue}\BibitemShut {NoStop}%
\bibitem [{\citenamefont {Bianco}\ and\ \citenamefont {Resta}(2011)}]{bianco}%
  \BibitemOpen
  \bibfield  {author} {\bibinfo {author} {\bibfnamefont {R.}~\bibnamefont {Bianco}}\ and\ \bibinfo {author} {\bibfnamefont {R.}~\bibnamefont {Resta}},\ }\bibfield  {title} {\bibinfo {title} {Mapping topological order in coordinate space},\ }\href {https://doi.org/10.1103/PhysRevB.84.241106} {\bibfield  {journal} {\bibinfo  {journal} {Phys. Rev. B}\ }\textbf {\bibinfo {volume} {84}},\ \bibinfo {pages} {241106} (\bibinfo {year} {2011})}\BibitemShut {NoStop}%
\bibitem [{\citenamefont {d'Ornellas}\ \emph {et~al.}(2022)\citenamefont {d'Ornellas}, \citenamefont {Barnett},\ and\ \citenamefont {Lee}}]{ornellas}%
  \BibitemOpen
  \bibfield  {author} {\bibinfo {author} {\bibfnamefont {P.}~\bibnamefont {d'Ornellas}}, \bibinfo {author} {\bibfnamefont {R.}~\bibnamefont {Barnett}},\ and\ \bibinfo {author} {\bibfnamefont {D.~K.~K.}\ \bibnamefont {Lee}},\ }\bibfield  {title} {\bibinfo {title} {Quantized bulk conductivity as a local chern marker},\ }\href {https://doi.org/10.1103/PhysRevB.106.155124} {\bibfield  {journal} {\bibinfo  {journal} {Phys. Rev. B}\ }\textbf {\bibinfo {volume} {106}},\ \bibinfo {pages} {155124} (\bibinfo {year} {2022})}\BibitemShut {NoStop}%
\bibitem [{\citenamefont {Sykes}\ and\ \citenamefont {Barnett}(2022)}]{Sykes_2022}%
  \BibitemOpen
  \bibfield  {author} {\bibinfo {author} {\bibfnamefont {J.}~\bibnamefont {Sykes}}\ and\ \bibinfo {author} {\bibfnamefont {R.}~\bibnamefont {Barnett}},\ }\bibfield  {title} {\bibinfo {title} {1d quasicrystals and topological markers},\ }\href {https://doi.org/10.1088/2633-4356/ac75a6} {\bibfield  {journal} {\bibinfo  {journal} {Materials for Quantum Technology}\ }\textbf {\bibinfo {volume} {2}},\ \bibinfo {pages} {025005} (\bibinfo {year} {2022})}\BibitemShut {NoStop}%
\bibitem [{\citenamefont {Sykes}\ and\ \citenamefont {Barnett}(2021)}]{sykes2}%
  \BibitemOpen
  \bibfield  {author} {\bibinfo {author} {\bibfnamefont {J.}~\bibnamefont {Sykes}}\ and\ \bibinfo {author} {\bibfnamefont {R.}~\bibnamefont {Barnett}},\ }\bibfield  {title} {\bibinfo {title} {Local topological markers in odd dimensions},\ }\href {https://doi.org/10.1103/PhysRevB.103.155134} {\bibfield  {journal} {\bibinfo  {journal} {Phys. Rev. B}\ }\textbf {\bibinfo {volume} {103}},\ \bibinfo {pages} {155134} (\bibinfo {year} {2021})}\BibitemShut {NoStop}%
\bibitem [{\citenamefont {Hannukainen}\ \emph {et~al.}(2022)\citenamefont {Hannukainen}, \citenamefont {Mart\'{\i}nez}, \citenamefont {Bardarson},\ and\ \citenamefont {Kvorning}}]{julia}%
  \BibitemOpen
  \bibfield  {author} {\bibinfo {author} {\bibfnamefont {J.~D.}\ \bibnamefont {Hannukainen}}, \bibinfo {author} {\bibfnamefont {M.~F.}\ \bibnamefont {Mart\'{\i}nez}}, \bibinfo {author} {\bibfnamefont {J.~H.}\ \bibnamefont {Bardarson}},\ and\ \bibinfo {author} {\bibfnamefont {T.~K.}\ \bibnamefont {Kvorning}},\ }\bibfield  {title} {\bibinfo {title} {Local topological markers in odd spatial dimensions and their application to amorphous topological matter},\ }\href {https://doi.org/10.1103/PhysRevLett.129.277601} {\bibfield  {journal} {\bibinfo  {journal} {Phys. Rev. Lett.}\ }\textbf {\bibinfo {volume} {129}},\ \bibinfo {pages} {277601} (\bibinfo {year} {2022})}\BibitemShut {NoStop}%
\bibitem [{\citenamefont {Fan}\ \emph {et~al.}(2023)\citenamefont {Fan}, \citenamefont {Zhang},\ and\ \citenamefont {Gu}}]{fan}%
  \BibitemOpen
  \bibfield  {author} {\bibinfo {author} {\bibfnamefont {R.}~\bibnamefont {Fan}}, \bibinfo {author} {\bibfnamefont {P.}~\bibnamefont {Zhang}},\ and\ \bibinfo {author} {\bibfnamefont {Y.}~\bibnamefont {Gu}},\ }\bibfield  {title} {\bibinfo {title} {{Generalized real-space Chern number formula and entanglement hamiltonian}},\ }\href {https://doi.org/10.21468/SciPostPhys.15.6.249} {\bibfield  {journal} {\bibinfo  {journal} {SciPost Phys.}\ }\textbf {\bibinfo {volume} {15}},\ \bibinfo {pages} {249} (\bibinfo {year} {2023})}\BibitemShut {NoStop}%
\bibitem [{\citenamefont {von Gersdorff}\ \emph {et~al.}(2021)\citenamefont {von Gersdorff}, \citenamefont {Panahiyan},\ and\ \citenamefont {Chen}}]{gersdorff}%
  \BibitemOpen
  \bibfield  {author} {\bibinfo {author} {\bibfnamefont {G.}~\bibnamefont {von Gersdorff}}, \bibinfo {author} {\bibfnamefont {S.}~\bibnamefont {Panahiyan}},\ and\ \bibinfo {author} {\bibfnamefont {W.}~\bibnamefont {Chen}},\ }\bibfield  {title} {\bibinfo {title} {Unification of topological invariants in dirac models},\ }\href {https://doi.org/10.1103/PhysRevB.103.245146} {\bibfield  {journal} {\bibinfo  {journal} {Phys. Rev. B}\ }\textbf {\bibinfo {volume} {103}},\ \bibinfo {pages} {245146} (\bibinfo {year} {2021})}\BibitemShut {NoStop}%
\bibitem [{\citenamefont {Ur\'{\i}a-\'Alvarez}\ \emph {et~al.}(2022)\citenamefont {Ur\'{\i}a-\'Alvarez}, \citenamefont {Molpeceres-Mingo},\ and\ \citenamefont {Palacios}}]{alvarez_2022}%
  \BibitemOpen
  \bibfield  {author} {\bibinfo {author} {\bibfnamefont {A.~J.}\ \bibnamefont {Ur\'{\i}a-\'Alvarez}}, \bibinfo {author} {\bibfnamefont {D.}~\bibnamefont {Molpeceres-Mingo}},\ and\ \bibinfo {author} {\bibfnamefont {J.~J.}\ \bibnamefont {Palacios}},\ }\bibfield  {title} {\bibinfo {title} {Deep learning for disordered topological insulators through their entanglement spectrum},\ }\href {https://doi.org/10.1103/PhysRevB.105.155128} {\bibfield  {journal} {\bibinfo  {journal} {Phys. Rev. B}\ }\textbf {\bibinfo {volume} {105}},\ \bibinfo {pages} {155128} (\bibinfo {year} {2022})}\BibitemShut {NoStop}%
\bibitem [{\citenamefont {Landauer}(1957)}]{landauer}%
  \BibitemOpen
  \bibfield  {author} {\bibinfo {author} {\bibfnamefont {R.}~\bibnamefont {Landauer}},\ }\bibfield  {title} {\bibinfo {title} {Spatial variation of currents and fields due to localized scatterers in metallic conduction},\ }\href {https://doi.org/10.1147/rd.13.0223} {\bibfield  {journal} {\bibinfo  {journal} {IBM Journal of Research and Development}\ }\textbf {\bibinfo {volume} {1}},\ \bibinfo {pages} {223} (\bibinfo {year} {1957})}\BibitemShut {NoStop}%
\bibitem [{\citenamefont {M.Buttiker}(1988)}]{buttiker}%
  \BibitemOpen
  \bibfield  {author} {\bibinfo {author} {\bibnamefont {M.Buttiker}},\ }\bibfield  {title} {\bibinfo {title} {Absence of backscattering in the quantum hall effect in multi-probe conductors},\ }\href@noop {} {\bibfield  {journal} {\bibinfo  {journal} {Phys. Rev. B}\ }\textbf {\bibinfo {volume} {38}},\ \bibinfo {pages} {9375} (\bibinfo {year} {1988})}\BibitemShut {NoStop}%
\bibitem [{\citenamefont {Groth}\ \emph {et~al.}(2014)\citenamefont {Groth}, \citenamefont {Wimmer}, \citenamefont {Akhmerov},\ and\ \citenamefont {Waintal}}]{groth1}%
  \BibitemOpen
  \bibfield  {author} {\bibinfo {author} {\bibfnamefont {C.~W.}\ \bibnamefont {Groth}}, \bibinfo {author} {\bibfnamefont {M.}~\bibnamefont {Wimmer}}, \bibinfo {author} {\bibfnamefont {A.~R.}\ \bibnamefont {Akhmerov}},\ and\ \bibinfo {author} {\bibfnamefont {X.}~\bibnamefont {Waintal}},\ }\bibfield  {title} {\bibinfo {title} {Kwant: a software package for quantum transport},\ }\href {https://doi.org/10.1088/1367-2630/16/6/063065} {\bibfield  {journal} {\bibinfo  {journal} {New Journal of Physics}\ }\textbf {\bibinfo {volume} {16}},\ \bibinfo {pages} {063065} (\bibinfo {year} {2014})}\BibitemShut {NoStop}%
\bibitem [{\citenamefont {Roy}\ and\ \citenamefont {Lu}(2025)}]{data}%
  \BibitemOpen
  \bibfield  {author} {\bibinfo {author} {\bibfnamefont {R.}~\bibnamefont {Roy}}\ and\ \bibinfo {author} {\bibfnamefont {Y.-M.}\ \bibnamefont {Lu}},\ }\href@noop {} {\bibinfo {title} {{Phase diagram of amorphous quantum spin Hall insulators}}} (\bibinfo {year} {2025}),\ \bibinfo {note} {\href{ https://doi.org/10.5281/zenodo.17544171}{https://doi.org/10.5281/zenodo.17544171}}\BibitemShut {NoStop}%
\end{thebibliography}
\end{document}